\documentclass[twocolumn,twocolappendix,numberedappendix]{openjournal}
\usepackage{amssymb}

\usepackage{color,comment}
\usepackage[normalem]{ulem}
\usepackage{amsmath}
\usepackage{ctable}
\usepackage{multirow}
\usepackage[breaklinks,colorlinks,citecolor=blue,urlcolor=blue,linkcolor=blue,filecolor=blue]{hyperref}

\let\oldAA=\AA
\renewcommand{\AA}{\mbox{\normalfont\oldAA}}

\newcommand{\qpah}[1]{$q_{\rm PAH}$ }
\newcommand{\lpah}[1]{$L_{\rm PAH}$ }
\newcommand{\lfir}[1]{$L_{\rm FIR}$ }
\newcommand{\Lpah}{L_{\rm PAH}}

\newcommand{\Mpah}{M_{\rm PAH}}

\newcommand{\Gzero}{G_0}
\newcommand{\Qabs}{Q_{\rm abs}}
\newcommand{\Kappa}{\kappa}

\newif\ifshowcode
\showcodefalse

\newcommand{\orcidauthor}[3]{\author{\href{http://orcid.org/#1}{#2$^{#3}$}}}

\shorttitle{The Lifecycle of PAHs in Cosmological Simulations}
\shortauthors{Narayanan et al.}

\begin{document}

\title{The Lifecycle and Emission Properties of PAHs in Cosmological Hydrodynamic Galaxy Formation Simulations\vspace{-1cm}}

\orcidauthor{0000-0002-7064-4309}{Desika Narayanan}{1,2}
\orcidauthor{0000-0002-5653-0786}{Paul Torrey}{3,4,5}
\orcidauthor{0000-0002-9729-3721}{Massimiliano Parente}{1}
\orcidauthor{0009-0001-6065-0414}{Grant P. Donnelly}{6}
\orcidauthor{0009-0008-7017-5742}{Dhruv T. Zimmerman}{1}
\orcidauthor{0000-0002-8111-9884}{Alex M. Garcia}{3,4,5}
\orcidauthor{0000-0001-6325-9317}{Helena M. Richie}{7}
\orcidauthor{0000-0003-1545-5078}{J.-D. T. Smith}{6}
\orcidauthor{0000-0001-7449-4638}{Brandon S. Hensley}{8}
\orcidauthor{0000-0003-3816-7028}{Federico Marinacci}{9,10}
\orcidauthor{0000-0002-6149-8178}{Jed McKinney}{11}
\orcidauthor{0000-0001-8592-2706}{Alexandra Pope}{12}
\orcidauthor{0000-0003-1151-4659}{Gerg\"o Popping}{13}
\orcidauthor{0000-0002-3790-720X}{Laura V. Sales}{14}
\orcidauthor{0000-0002-4378-8534}{Karin Sandstrom}{15}
\orcidauthor{0000-0002-2919-1109}{Ethan Savitch}{1}
\orcidauthor{0000-0003-4702-7561}{Irene Shivaei}{16}
\orcidauthor{0000-0003-3256-5615}{Justin Spilker}{17}
\orcidauthor{0000-0003-2093-4452}{Cory M. Whitcomb}{6}

\affil{$^{1}$Department of Astronomy, University of Florida, 211 Bryant Space Sciences Center, Gainesville, FL 32611 USA \\
$^{2}$Cosmic Dawn Center at the Niels Bohr Institute, University of Copenhagen and DTU-Space, Technical University of Denmark \\
$^{3}$Department of Astronomy, University of Virginia, 530 McCormick Road, Charlottesville, VA 22903, USA \\
$^{4}$Virginia Institute for Theoretical Astronomy, University of Virginia, Charlottesville, VA 22904, USA \\
$^{5}$The NSF-Simons AI Institute for Cosmic Origins, USA \\
$^{6}$Ritter Astrophysical Research Center, Department of Physics and Astronomy, University of Toledo, Toledo, OH 43606, USA \\
$^{7}$Physics and Astronomy Department, University of Pittsburgh, 3941 O'Hara St. Pittsburgh, PA, 15260, USA \\
$^{8}$Jet Propulsion Laboratory, California Institute of Technology, 4800 Oak Grove Drive, Pasadena, CA 91109, USA \\
$^{9}$Department of Physics and Astronomy ``Augusto Righi'', University of Bologna, via Gobetti 93/2, 40129, Bologna, Italy \\
$^{10}$INAF, Astrophysics and Space Science Observatory Bologna, Via P. Gobetti 93/3, 40129 Bologna, Italy \\
$^{11}$Department of Astronomy, University of Texas at Austin, Austin, TX, USA \\
$^{12}$Department of Astronomy, University of Massachusetts, Amherst, MA, 01003, USA \\
$^{13}$European Southern Observatory, Karl-Schwarzschild-Str. 2, 85748, Garching, Germany \\
$^{14}$Department of Physics and Astronomy, University of California, Riverside, CA 92507, USA \\
$^{15}$Department of Astronomy \& Astrophysics, University of California, San Diego, 9500 Gilman Drive, La Jolla, CA 92093 \\
$^{16}$Centro de Astrobiolog\'{i}a (CAB), CSIC-INTA, Carretera de Ajalvir km 4, Torrej\'{o}n de Ardoz, 28850, Madrid, Spain \\
$^{17}$Department of Physics and Astronomy and George P. and Cynthia Woods Mitchell Institute for Fundamental Physics and Astronomy, Texas A\&M University, 424 TAMU, College Station, TX, 77843, USA}
\thanks{E-mail: \href{mailto:desika.narayanan@ufl.edu}{desika.narayanan@ufl.edu}}

\begin{abstract}
We present the first cosmological model for the lifecycle and luminous
properties of polycyclic aromatic hydrocarbons (PAHs) in galaxies as
they evolve from $z=6\rightarrow 0$.  We model $40$ galaxies with the
cosmological hydrodynamic zoom-in technique, coupled with an
on-the-fly model for the formation, growth and destruction of
multi-species, multi-size dust grains in the ISM.  We approximate PAHs as ultrasmall ($a < 13$~\AA) carbonaceous dust
grains, and couple this model with single-photon excitation
calculations to compute the emergent mid-infrared spectra.  Our main
results follow.  (1) If we assume that dust is large upon formation,
then PAHs are naturally able to form {\it in situ} in the ISM via
grain-grain shattering.  Interstellar collision velocities increase in
low density, diffuse gas in our model; as galaxies evolve, the
increase in fractional mass of diffuse gas drives an increase in
grain-grain collision velocities and a corresponding rise in the PAH
mass fraction ($q_{\rm PAH}$) from $\sim 5 \times 10^{-4}$ at $z\sim
4$ to $\sim 10^{-2}$ at $z\sim 0$.  (2) Increased PAH production in
the diffuse ISM results in an inverse relationship between the PAH
mass fraction ($q_{\rm PAH}$) and the molecular gas fraction.  (3) The
PAH light-to-mass ratio scales linearly with the radiation field
intensity ($L_{\rm PAH}/M_{\rm PAH} \propto G_0$) but anti-correlates
with $q_{\rm PAH}$, because high-$\Sigma_{\rm SFR}$ galaxies have a
denser ISM that suppresses shattering.  This decoupling means that the
physical $q_{\rm PAH}$ and the observed $L_{\rm PAH}/L_{\rm FIR}$ do
not evolve in lockstep.  (4) The PAH-metallicity relationship (PZR)
arises naturally in this framework: galaxies enrich and grow their
diffuse ISM fraction simultaneously, linking rising metallicity to
rising $q_{\rm PAH}$.  Our models represent the first to reproduce the
PZR observed across $z=0$-$2$.  (5) A relationship between both SFR
  and $L_{\rm PAH}$ as well as $M_{\rm mol}$ and $L_{\rm PAH}$ emerges from
two effects: more massive galaxies have larger PAH reservoirs, and
higher-SFR galaxies excite their PAHs more efficiently per unit mass.
Taken together, these results suggest that grain-grain shattering
  in the diffuse ISM is the main driver behind the evolution of cosmic
  PAH abundances, providing a unified physical framework that
  simultaneously accounts for the PZR, $L_{\rm PAH}$-SFR, and $L_{\rm
    PAH}$-$M_{\rm mol}$ relations, and the redshift evolution of PAH
  emission in galaxies.
\end{abstract}



\section{Introduction}
\label{section:introduction}
The mid-infrared (MIR) spectral energy distributions (SEDs) of
star-forming galaxies are dominated by a series of strong emission
features between $\lambda \approx 3.3-17 \mu$m.  First detected in the spectra of gas surrounding ionized nebulae by
\citet{gillett73a} and \citet{merrill75a}, these features were
subsequently attributed to the vibrational de-excitation of polycyclic
aromatic hydrocarbon (PAH) molecules.  These large carbonaceous species
contain tens to hundreds of carbon atoms arranged in aromatic
rings, and are stochastically heated to $T \gtrsim 10^3$~K by the
absorption of single ultraviolet (UV) photons
\citep{leger84a,allamandola85a,puget89a}.

PAH emission is nearly ubiquitous in star-forming galaxies in the low-redshift Universe
\citep[e.g.,][]{genzel98a,helou00a,smith07a}, and the integrated
luminosity across all PAH bands can readily constitute 10-20\% of the
total infrared luminosity of a galaxy
\citep[e.g.,][]{helou00a,smith07a,dale09a,lai20a}.  Physically, PAHs
may be an important catalyst for molecular hydrogen formation on grain
surfaces \citep{thrower12a,foley18a,rodriguezmontero26a}, as well as
the dominant source of photoelectric heating in the neutral phase of
the ISM \citep{bakes94a}.  Ultimately, the luminosity of PAHs in
galaxies is set by (i) the abundance of ultrasmall carbonaceous dust
grains and (ii) the availability of UV photons to excite them
\citep[see][for a relatively recent review]{li20b}.

Despite their outsized role in both the energetics and chemistry of the ISM, the physical processes that govern the lifecycle of PAHs across cosmic time
  remain poorly understood.  Are PAHs formed via
injection into the ISM from evolved stars, or do they grow {\it in
  situ} via either the growth of molecules, or shattering of larger
dust grains?  What physical conditions set the mass fraction of PAHs
over cosmic time?  What drives the major observed scaling relations,
and, relatedly, does the luminosity from PAHs primarily trace $M_{\rm
  PAH}$, or the ambient exciting ultraviolet radiation field?  The
questions surrounding PAHs are still fundamental in nature.

The {\it Spitzer Space Telescope} uncovered a number of key empirical
scaling relations that offer important clues to these questions.  The
first is that PAH luminosity correlates approximately linearly with
the global star formation rate (SFR) of galaxies.  For example,
\citet{peeters04a} showed that the 6.2~$\mu$m PAH feature tracks the
far infrared (FIR) luminosity in star-forming regions and starburst
galaxies, while \citet{wu05b} demonstrated that 8~$\mu$m luminosity
(dominated by the 7.7~$\mu$m PAH complex) traces both H$\alpha$ and
1.4~GHz radio luminosities over $\sim$2-3 orders of magnitude.
Subsequent spectroscopic calibrations established the 6.2, 7.7, and
11.3~$\mu$m features individually as quantitative SFR indicators
\citep{shipley16a,maragkoudakis18a,xie19a}.  At high-redshift,
\citet{pope08a} used the 6.2~$\mu$m feature to establish a linear
$L_{\rm PAH}$-$L_{\rm FIR}$ relation for submillimeter galaxies out
to $z \sim 2$.  This said, the nature of the PAH-SFR relationship is
not fully understood.  Whether such a relationship exists because
``bigger things are bigger'' (i.e., higher SFR typically correlates
with higher galaxy mass, and hence, higher $M_{\rm PAH}$), or because
higher SFR galaxies have a larger UV radiation field is unclear.
\citet{bendo08a} suggested that the 8~$\mu$m emission on resolved
scales may even trace the diffuse radiation field more closely than
active star-forming regions.  Alternatively, it is possible that the
$L_{\rm PAH}$-SFR relationship more fundamentally traces a $M_{\rm
  PAH}$-molecular gas mass ($M_{\rm mol}$) relationship \citep[e.g.][]{whitcomb23a,chown21a,zhang23a},
and that the observed $L_{\rm PAH}$-SFR relationship is actually just
a manifestation of the \citet{kennicutt98a} star formation relation.
Despite these caveats, PAH-based SFR measurements have been extended
to high redshift, with the rest-frame MIR used to infer total infrared
luminosities and SFRs for galaxies at $z \sim 1$-3
\citep{reddy06a,yan07a,siana09a,wuyts11a,pope08a,rujopakarn13a}.

The second key galaxy-wide PAH scaling relation is the PAH-metallicity
relation \citep[following the nomenclature of ][hereafter, the
  PZR]{whitcomb24a}.  There is now significant observational evidence
that the PAH luminosity fraction ($L_{\rm PAH}/L_{\rm FIR}$) and, by
inference, the PAH mass fraction ($q_{\rm PAH} \equiv M_{\rm
  PAH}/M_{\rm dust}$) are suppressed at subsolar metallicities
\citep[e.g.,][]{engelbracht05a,jackson06a,madden06a,draine07a,smith07a,gordon08a,munoz-mateos09a,hunt10a,sandstrom12a,aniano20a,lai20a,whitcomb20a,shivaei24a}.
This said, despite widespread evidence for the PZR, its physical
origin remains debated.  The most commonly invoked mechanism is the
photodestruction of PAHs in hard, poorly shielded radiation fields
\citep{voit92a,madden06a,gordon08a,engelbracht08a,egorov23a}, though
other proposed mechanisms include increased erosion via thermal
sputtering in a hotter ISM \citep{hunt11a}, a lack of seed metals to
grow into small grains and/or a lack of seed dust grains that shatter
into smaller particles \citep{seok14a}, and delayed injection by
carbon-rich AGB stars in young, low-metallicity systems that have not
yet had time to build up a significant PAH reservoir
\citep{galliano08a}.  More recently, \citet{whitcomb25a} and
\citet{tarantino25a} have argued that the PZR may be driven primarily
by inhibited growth of PAH grains via accretion of carbon atoms and
other PAH precursors in the dense ISM, a process that becomes
increasingly inefficient as gas-phase metallicity decreases.
Alternatively, some studies have argued that the primary
anti-correlation with PAH mass fraction and metallicity is not due to
metallicity itself, but rather the radiation field hardness
\citep{gordon08a}.  Regardless of the physical origin, the observed
PZR offers an important constraint on the lifecycle of PAHs in
galaxies over cosmic time.

The James Webb Space Telescope (JWST) has expanded our understanding
of the PAH lifecycle in two key ways: by extending our understanding
of the aforementioned global scaling relations to high-redshift, and
by providing a relatively high-resolution view of PAHs in the ISM of
local galaxies.  For example, \citet{shivaei24a} analyzed MIRI
observations of hundreds of galaxies between $z=0.7-2$, and found a
galaxy-integrated PZR relation comparable to that inferred in the
local Universe.  At the same time, \citet{spilker23a} confirmed the
existence of PAHs (via the $3.3 \mu$m feature) at $z>4$, while
\citet{markov25a}, \citet{ormerod25a} and \citet{witstok23a} all inferred the
presence of the $2175 \AA$ attenuation bump at $z>6$, which, if
related to PAHs \citep{sandstrom10a,shivaei22a} suggests ultrasmall
carbonaceous grains within the first billion years of the Universe.
\citet{mckinney25a} confirmed the relationship between $L_{\rm PAH}$
and SFR at $z\sim 1-2$, relating the $3.3 \mu$m feature to the
bolometric infrared luminosity.

Alongside expanding PAH studies to the high-redshift Universe, JWST has enabled resolved observations of local
galaxies that have provided an unprecedented view of PAH emission
across the multiphase ISM of nearby galaxies.  For example,
\citet{chastenet23a,sutter24a} find that outside of H~{\sc ii}
regions, the PAH fraction of nearby galaxies \citep[observed as a part
  of the PHANGS-JWST survey;][]{lee23a} is remarkably constant, with
$q_{\rm PAH} \sim 3$-$6\%$ across both diffuse and molecular gas,
suggesting that PAHs are relatively well-mixed with the parent dust
population.  At the same time, \citet{leroy23a} showed that the
11.3~$\mu$m PAH band scales approximately linearly with gas surface
density down to $\Sigma_{\rm gas} \sim 1 M_\odot {\rm pc}^{-2}$.
Within H~{\sc ii} regions, however, the picture changes dramatically:
\citet{sandstrom10a}, \citet{egorov23a}, \citet{sutter24a} and \citet{murgia26a} all
found a strong anti-correlation between PAH fraction and ionization
parameter for $\sim$1500 H~{\sc ii} regions, with the relation
steepening for the most luminous regions, suggesting PAHs may be
destroyed in ionized gas.  On sub-kpc scales, \qpah \ decreases
steeply with increasing star formation rate surface density
($\Sigma_{\rm SFR}$) and specific star formation rate (sSFR)
\citep{chastenet25a}.  These results provide important
  clues for what drives the relationship between the PAH abundance in
galaxies, the phase structure of the ISM, and the galaxy assembly
  history itself.

The goal of this paper is to contextualize these observed scaling
relations within a broader theory for the lifecycle of PAHs in the ISM
of cosmologically evolving galaxies.  While there is indeed a rich
history of numerically modeling the evolution of dust masses and grain
sizes in galaxies, to date there are no {\it cosmological} models for
the evolution of PAHs in galaxies that we are aware of.  Existing
approaches broadly fall into three categories of increasing
complexity.  The simplest track total dust mass with a fixed grain
size distribution
\citep[e.g.,][]{dwek98a,zhukovska08a,mckinnon16a,mckinnon17a,popping17a,li19a,parente22a,choban25a}.
These models capture the dust mass budget but cannot predict PAH
abundances, as they do not model the small-grain population.  A
second generation of models adopts the two-size approximation of
\citet{hirashita15a}, splitting grains into ``small'' ($a < 0.03 \;
\mu$m) and ``large'' ($a > 0.03 \; \mu$m) populations that each
evolve via accretion, shattering, coagulation, and supernova
destruction.  This framework has been widely implemented in
hydrodynamic simulations
\citep[e.g.,][]{aoyama17a,aoyama18a,gjergo18a,hou19a,granato21a,parente22a,dubois24a,trayford26a},
but the small-grain bin serves only as a proxy for PAHs and the
boundary between bins is a tunable choice.  The most physically
rigorous approach discretizes the full grain size distribution into
$\sim$16-32 logarithmic bins, enabling explicit tracking of the
ultrasmall grain population
\citep[e.g.,][]{asano13b,mckinnon18a,aoyama20a,choban26a,parente26a}.  One-zone and
semi-analytic models in this category have demonstrated that PAH
abundances can be understood through the interplay of shattering,
accretion, and aromatization
\citep{seok14a,rau19a,hirashita20a,hirashita20b,hirashita22a}, while
\citet{narayanan23a} embedded a 16-bin grain size distribution within
3D hydrodynamic simulations of idealized galaxies, coupling on-the-fly
dust evolution to radiative transfer for self-consistent PAH emission
predictions.  Most recently, \citet{rodriguezmontero26a} developed an
on-the-fly dust and PAH evolution module for the {\sc ramses}
radiation-hydrodynamics code, treating PAHs as separate species with
their own formation and destruction channels, though this has so far
been applied only to isolated galaxies.


In this paper, we present the first ever cosmological zoom-in simulations
developed to model both the physical, and luminous properties of PAHs
in galaxies as they evolve across cosmic time.  Our main goal is to
understand both the lifecycle of PAHs, as well as how they relate to
global galaxy scaling relations (we defer studies of the resolved
properties of PAHs to a future paper).  At the outset, we
  emphasize that while PAHs are most likely in reality molecules
  \citep[e.g.][]{tielens08a}, our model treats their physical
  evolution, as an approximation, as the smallest carbonaceous grains
  in an evolving grain size distribution.  This said, the emission
  from PAHs is computed through single photon stochastic heating of
  molecules.  We describe this approximation and its limitations in
  \S~\ref{section:methods} and \S~\ref{section:caveats}.  Our paper
is organized as follows. In \S\ref{section:methods} we describe the
simulation methodology, including the cosmological zoom-in framework,
the on-the-fly dust and PAH evolution model, and the radiative
transfer post-processing used to generate synthetic MIR SEDs.  In
\S\ref{section:lifecycle} we present a model for the {\it physical}
lifecycle of PAHs in our simulation, while in
\S~\ref{section:luminous_properties}, we follow with how these
translate to the {\it luminous} properties of PAHs in galaxies as they
evolve across cosmic time.  Here, we focus specifically on the
application of our model to the PZR, the SFR-$L_{\rm PAH}$, and the
$L_{\rm PAH}$-$M_{\rm mol}$ relationships in galaxies.  We discuss our
results in the context of other theoretical models and observations in
\S\ref{section:discussion}, and summarize our conclusions in
\S\ref{section:summary}.

\section{Simulation Methodology}
\label{section:methods}

\subsection{Cosmological Galaxy Formation Simulations}
\label{section:methods_sims}
We explore the lifecycle and luminous properties of PAHs across cosmic
time by employing cosmological hydrodynamic zoom-in galaxy formation
simulations, run with the moving-mesh hydrodynamics and $N$-body code
{\sc arepo} \citep{springel10a,weinberger20a}.  Within {\sc arepo}, we
employ the {\sc smuggle} galaxy formation physics model
\citep{marinacci19a}.  We briefly describe {\sc smuggle} here, though refer to \citet{marinacci19a} for more details.

The ISM physics includes radiative cooling from primordial species
(via two-body collisions, free-free emission, recombination, and
Compton cooling off the CMB) and metal-enriched gas (via line cooling
rates computed from {\sc cloudy} photoionization models;
\citealt{ferland13a,vogelsberger13a}).  At low temperatures ($T \sim
10$--$10^4$ K), we supplement these cooling rates with molecular and
fine-structure emission rates drawn from the fitting functions of
\citet{hopkins18a}.  We implement gas self-shielding from the metagalactic UV
background above densities $n > 10^{-3}$ cm$^{-3}$,
following the redshift-dependent parameterization of
\citet{rahmati13a}.  We include heating from cosmic rays via
the density-dependent prescription of \citet{guo08b}, and
photoelectric heating following \citet{wolfire03a}.

Star formation proceeds in gravitationally bound molecular gas at
densities above a threshold density $n_{\rm thresh} \sim 500$ cm$^{-3}$, where the star
formation rate follows a volumetric star formation relation
\citep{kennicutt98a}:
\begin{equation}
  \dot{M}_* = \epsilon_{\rm ff} \frac{M_{\rm gas}}{t_{\rm ff}}
  \label{equation:sfr}
\end{equation}
where $t_{\rm ff} = \sqrt{3\pi/(32 G \rho_{\rm gas})}$ is the local
free-fall time and $\epsilon_{\rm ff}$ is the star formation
efficiency per free-fall time \citep{krumholz07a}.  We assume
$\epsilon_{\rm ff} = 0.01$.  We compute the molecular gas fraction
following the \citet{krumholz08a,krumholz09a} prescription, which ties
the H$_2$ fraction to gas surface density and metallicity.  {\sc smuggle} models the production and advection of nine individual metal species.
Stellar feedback includes energy and momentum injection from Type Ia
and Type II supernovae (assuming a \citealt{chabrier03a} IMF with a
delay-time distribution for Type Ia events), photoionization,
radiation pressure, and mass loss from OB and AGB stellar winds.  

We generate the initial conditions for our zoom-in simulations with
{\sc music} \citep{hahn11a}.  We first perform a
low-resolution dark matter-only simulation in a $(100\;h^{-1}\;{\rm
  Mpc})^3$ cosmological box with cosmological parameters $\Omega_m =
0.3$, $\Omega_\Lambda = 0.7$, and $h = 0.7$, initialized at $z=99$
and evolved to $z=0$.  We then identify target halos with the {\sc caesar} group finder\footnote{\url{https://caesar.readthedocs.io}}, and trace all particles within
$2.5\times$ the radius of the most distant dark matter particle back to the initial conditions to form the high-resolution
mask.  We resimulate the zoom regions at high
resolution with full hydrodynamics, the {\sc smuggle} physics model,
and the on-the-fly dust evolution model described in
\S~\ref{section:methods_dust}.  The resulting baryon mass resolution is
$\sim 8.8 \times 10^4/h$ M$_\odot$.

Our simulation sample consists of $40$ zoom-in galaxies, spanning a
broad range of stellar masses ($\log(M_*/{\rm M}_\odot) \approx
8.2$--$10.9$), halo masses ($\log(M_{200}/{\rm M}_\odot) \approx
10.9$--$12.7$), and gas-phase metallicities ($12 + \log({\rm O/H})
\approx 6$--$9.7$) at their final snapshots.  The simulations are
evolved to a wide range of redshifts: of the $40$ galaxies, $8$ are
evolved to $z=0$, $13$ to $z < 1$, $21$ to $z < 2$, and all $40$ to at
least $z \approx 5.5$.  The final redshifts are determined by the
  feasibility of simulation completion given the time steps.
  Astrophysical processes, such as mergers or the formation of
  extremely dense clumps can cause the time steps to become extremely
  short, making further progress in some of the simulations prohibitive. We
summarize the physical properties of all model galaxies at their final
snapshots in Table~\ref{table:galaxy_sample}.

We note that we do not include black holes in our simulations.
  Black holes of course have the ability to impact the galaxy physical
  properties, as well as potentially the emission from PAHs.  We are
  exploring some aspects of the impact of AGN emission on the mid-IR
  spectrum of galaxies in G. Donnelly et al. (in prep).

Finally, we note a point of difference in the models run here
  compared to similar (aside from the mass range and redshift range
  probed) models presented in \citet{narayanan26a}.  In the current
  presented models, we have adjusted $3$ feedback parameters in order
  to inhibit the formation of overly dense bulges, which prevents the
  simulations from proceeding to $z=0$, owing to the aforementioned time-stepping issues.
  In particular, we increase the thermal feedback efficiency (a
  dimensionless efficiency factor applied to the thermal energy
  injected into gas cells by stellar feedback) from $\sim
  10$ to $\sim 20$, decrease the stellar age up to which radiation momentum
  feedback is injected by young star particles (from $10$ Myr to $3$
  Myr), and the luminosity-to-mass ratio assigned to young star
  particles for the purpose of radiation feedback (from $\sim 200$ to
  $\sim 5000$).  We note that while we adjusted these parameters to
  ensure reasonable disk formation (i.e., without forming dense clumps
  that create galaxies that exceed \citet{tully77a} constraints), we
  have not tuned to any PAH properties.  In the Appendix
  (\S~\ref{appendix:galaxy_physical_properties}), we demonstrate model
  viability via the simulated $M_*-M_{\rm halo}$ relationship and
  $M_*-Z_{\rm gas}$ relationship for our model galaxies from $z=0-6$.  We defer a full exploration of the sensitivity of the {\sc smuggle} model to feedback parameters to a future study.

\begin{deluxetable*}{lccccccc}
  \tablecaption{Physical properties of our model galaxy sample,
    measured at each galaxy's final simulation snapshot.  Columns are: galaxy
    name, $z_{\rm final}$, $M_*$, $M_{\rm gas}$, $M_{\rm dust}$, $M_{\rm 200}$, metallicity, and SFR.\label{table:galaxy_sample}}
  \tablehead{
    \colhead{Galaxy} &
    \colhead{$z_{\rm final}$} &
    \colhead{$\log \frac{M_*}{{\rm M}_\odot}$} &
    \colhead{$\log \frac{M_{\rm gas}}{{\rm M}_\odot}$} &
    \colhead{$\log \frac{M_{\rm dust}}{{\rm M}_\odot}$} &
    \colhead{$\log \frac{M_{200}}{{\rm M}_\odot}$} &
    \colhead{$12+\log({\rm O/H})$} &
    \colhead{SFR} \\
    \colhead{} & \colhead{} & \colhead{} & \colhead{} & \colhead{} & \colhead{} & \colhead{} &
    \colhead{(M$_\odot$ yr$^{-1}$)}
  }
  \startdata
  294    & 5.49 &  9.71 & 10.39 & 8.92 & 11.49 & 9.70 &  35.5 \\
  525    & 3.74 & 10.22 & 10.77 & 8.15 & 11.91 & 8.80 &  30.6 \\
  546    & 3.26 & 10.12 & 10.64 & 8.17 & 11.83 & 8.74 &  19.6 \\
  655    & 0.75 & 10.44 & 11.26 & 7.17 & 12.53 & 8.59 &   2.7 \\
  820    & 0.50 & 10.75 & 11.28 & 6.39 & 12.68 & 9.22 &  22.7 \\
  867    & 2.00 &  9.91 & 10.45 & 6.96 & 11.66 & 8.36 &   6.5 \\
  983    & 2.00 &  9.91 & 10.39 & 7.05 & 11.74 & 8.10 &  12.7 \\
  1025   & 2.51 & 10.87 & 10.90 & 9.35 & 12.22 & 9.26 & 431.9 \\
  1101   & 2.00 & 10.43 & 10.63 & 8.78 & 12.01 & 9.09 &  10.9 \\
  1217   & 1.75 & 10.28 & 10.05 & 7.38 & 11.72 & 8.61 &   9.3 \\
  1281   & 2.25 & 10.47 & 10.92 & 9.14 & 12.10 & 9.63 &  13.2 \\
  1446   & 0.00 & 10.68 & 10.98 & 6.64 & 12.54 & 9.28 &   5.0 \\
  1480   & 2.25 & 10.45 & 10.75 & 8.74 & 12.11 & 8.91 &  10.3 \\
  2096   & 2.51 & 10.20 & 10.54 & 7.15 & 11.90 & 8.25 &   1.5 \\
  2156   & 3.74 &  9.52 & 10.16 & 6.46 & 11.24 & 8.35 &  24.6 \\
  2625   & 2.25 & 10.48 & 10.54 & 8.63 & 11.92 & 8.74 &  25.8 \\
  3103   & 2.00 & 10.27 & 10.26 & 7.84 & 11.55 & 8.52 &   9.0 \\
  3535   & 2.00 & 10.17 & 10.53 & 7.33 & 11.80 & 8.36 &  16.0 \\
  4091   & 0.00 &  9.79 & 10.58 & 6.57 & 11.83 & 8.83 &   2.0 \\
  4120   & 1.00 & 10.21 & 10.00 & 7.01 & 11.56 & 8.12 &   2.5 \\
  5383   & 2.51 &  9.46 &  9.70 & 6.50 & 11.15 & 8.35 &   2.3 \\
  5462   & 0.00 &  9.55 & 10.43 & 6.96 & 11.80 & 7.37 &   0.1 \\
  5481   & 1.00 &  9.82 & 10.31 & 6.71 & 11.43 & 8.16 &   4.0 \\
  5566   & 2.51 &  9.69 & 10.13 & 6.70 & 11.38 & 8.28 &   5.3 \\
  6105   & 4.00 &  9.91 & 10.50 & 6.95 & 11.71 & 8.21 &  66.7 \\
  6120   & 3.00 &  9.85 & 11.05 & 7.11 & 12.05 & 7.33 &   6.9 \\
  6266   & 0.00 & 10.02 &  9.90 & 6.74 & 11.76 & 8.39 &   0.5 \\
  6351   & 1.00 &  9.83 &  9.95 & 6.93 & 11.45 & 7.64 &   0.4 \\
  7126   & 0.75 &  9.53 &  9.70 & 6.15 & 10.91 & 7.83 &   0.7 \\
  7801   & 0.00 &  8.16 & 10.77 & 6.14 & 11.52 & 6.08 &   0.0 \\
  8047   & 0.25 &  9.98 &  9.83 & 6.42 & 11.48 & 8.29 &  21.3 \\
  8750   & 1.50 &  9.89 & 10.05 & 6.93 & 11.29 & 8.39 &   2.1 \\
  8752   & 0.25 &  9.79 &  9.72 & 6.97 & 11.33 & 8.05 &   0.2 \\
  9165   & 0.00 & 10.16 & 10.20 & 6.55 & 11.61 & 7.85 &   5.1 \\
  9265   & 0.75 &  9.51 & 10.10 & 5.56 & 11.54 & 7.93 &   0.1 \\
  9283   & 0.50 &  9.85 &  9.87 & 6.77 & 11.27 & 8.22 &   0.6 \\
  9340   & 0.50 &  9.99 &  9.79 & 5.90 & 11.50 & 6.86 &   1.5 \\
  10212  & 0.00 &  9.66 &  9.64 & 6.90 & 11.53 & 8.04 &   0.1 \\
  10407  & 0.00 &  9.92 &  9.36 & 6.69 & 11.49 & 8.26 &   1.0 \\
  10813  & 1.50 & 10.03 &  9.91 & 7.08 & 11.44 & 8.39 &   4.3 \\
  \enddata
\end{deluxetable*}

\subsection{Dust Modeling}
\label{section:methods_dust}

We model the on-the-fly production, destruction, and evolution of dust
grains directly within our hydrodynamic simulations, building on the
framework of \citet{mckinnon16a,mckinnon17a,mckinnon18a}, with
significant updates from \citet{li21a} and \citet{narayanan23a}.
Here, we outline the essential elements of the dust model as
implemented into {\sc smuggle}/{\sc arepo}, and refer to
\citet{narayanan23a} for a full description of the underlying dust
algorithms, including the PAH model presented here.

{\sc arepo} is a moving-mesh finite-volume code in which gas is
  evolved on an unstructured Voronoi mesh.  In our implementation,
  dust is not treated as a passive scalar subcomponent of the gas
  cells, but instead as a distinct population of Lagrangian
  super-particles, following the life dust particle approach of
  \citet{bekki15a,mckinnon18a} and \citet{li21a}.  Here, each dust
  super-particle represents an ensemble of physical grains, and
  carries its own grain size distribution, discretized into $16$
  logarithmic bins for both carbonaceous and silicate species
  (\S~\ref{section:dust_production}).  Dust particles are spawned
  stochastically from star particles as evolved stars return condensed
  metals, are dynamically evolved subject to gravity and aerodynamic
  drag against the local gas flow, and are split or merged to maintain
  mass resolution \citep{li21a}.  Mass exchange with the gas is
  handled locally: grain growth via accretion draws from the gas-phase
  metal content of neighboring cells, while thermal sputtering and
  shock destruction return dust mass to the metal fields of the local
  gas cells (\S~\ref{section:dust_growth_destruction}).  This
  live-particle approach differs from implementations in which dust is
  carried as a scalar property advected with the gas fluid elements
  \citep[e.g.][]{aoyama17a,mckinnon16a,mckinnon17a,li19a,graziani20a,kannan22a},
  and allows the dust to dynamically decouple from the gas while
  carrying a spatially varying grain size distribution.

\subsubsection{Dust Production}
\label{section:dust_production}

Dust originates through the condensation of metals returned to the
ISM by evolved stars \citep{dwek98a}.  We produce dust particles
directly from star particles, adopting dust yields from
\citet{schneider14a} for AGB stars and from \citet{nozawa10a} for
Type II supernovae.  We track dust as two chemical
species, silicates and carbonaceous grains, with internal grain
densities of $3.3$ g cm$^{-3}$ and $2.2$ g cm$^{-3}$, respectively.
For a given dust particle, the total carbon mass corresponds to the
carbonaceous component and the remainder to the silicate component.

The carbon-to-oxygen ratio (C/O) of the stellar ejecta controls
which dust species forms \citep{dwek98a,ferrarotti06a}.  Because CO
molecule formation is thermodynamically favored, CO locks up
whichever of carbon or oxygen is less abundant, and only the 
  excess of the dominant element condenses into grains.  For
carbon-rich AGB stars (C/O $> 1$),  the dust mass of species $i$
is:
\begin{equation}
  m_{i,d}^{\rm AGB} = \begin{cases}
    \delta_{\rm C}^{\rm AGB}\left(m_{\rm C,ej}^{\rm AGB} - 0.75\, m_{\rm O,ej}^{\rm AGB}\right), & i = {\rm C} \\
    0, & {\rm otherwise}
  \end{cases}
  \label{equation:agb_co_gt1}
\end{equation}
where $m_{i,d}^{\rm AGB}$ is the mass of dust species $i$ produced per AGB star, and $m_{\rm C,ej}^{\rm AGB}$ and $m_{\rm O,ej}^{\rm AGB}$ are the masses of carbon and oxygen ejected by the star. The factor $0.75\,m_{\rm O,ej}^{\rm AGB}$ accounts for the
carbon mass locked in CO, and $\delta_i^{\rm AGB}$ is the
condensation efficiency.  For oxygen-rich AGB stars (C/O $< 1$), we assume that all
carbon is bound in CO and the remaining oxygen combines with
refractory metals to form silicate dust:
\begin{equation}
  m_{i,d}^{\rm AGB} = \begin{cases}
    0, & i = {\rm C} \\
    16 \displaystyle\sum_{j={\rm Mg,Si,S,Ca,Fe}} \delta_j^{\rm AGB}\, m_{j,\rm ej}^{\rm AGB}/\mu_j, & i = {\rm O} \\
    \delta_i^{\rm AGB}\, m_{i,\rm ej}^{\rm AGB}, & {\rm otherwise.}
  \end{cases}
  \label{equation:agb_co_lt1}
\end{equation}
Here, $\mu_j$ is the atomic mass of element $j$ (in atomic mass units), and $m_{j,\rm ej}^{\rm AGB}$ is the mass of species $j$ ejected by the star.
Type II supernovae, in contrast, produce both dust species
simultaneously because their ejecta are macroscopically mixed, with
condensation proceeding independently in carbon-rich and oxygen-rich
zones \citep{nozawa10a,dwek98a}:
\begin{equation}
  m_{i,d}^{\rm SNII} = \begin{cases}
    \delta_{\rm C}^{\rm SNII}\, m_{\rm C,ej}^{\rm SNII}, & i = {\rm C} \\
    16 \displaystyle\sum_{j={\rm Mg,Si,S,Ca,Fe}} \delta_j^{\rm SNII}\, m_{j,\rm ej}^{\rm SNII}/\mu_j, & i = {\rm O} \\
    \delta_i^{\rm SNII}\, m_{i,\rm ej}^{\rm SNII}, & {\rm otherwise}
  \end{cases}
  \label{equation:snii_dust}
\end{equation}
We adopt fixed condensation efficiencies of $\delta_i^{\rm AGB} =
0.2$ and $\delta_i^{\rm SNII} = 0.15$
\citep{bianchi07a,ferrarotti06a}.  As a result, the carbonaceous
fraction of dust in our simulations grows with cosmic time as low-
and intermediate-mass stars reach the AGB phase.

A key assumption of our model is that stellar dust is large upon
  formation.  We do not explicitly simulate the destruction of grains
  by reverse shocks in supernova remnants; instead, we adopt initial
  grain size distributions for newly formed stardust that implicitly
  account for this processing.  In this formalism, small grains
  produced in supernovae are expected to be efficiently destroyed in
  reverse shocks as found in a number of theoretical studies
  \citep{bianchi07a,nozawa07a,silvia12a,otaki26b,otaki26a}; consistent
  with this, observations of dust in core-collapse supernova ejecta
  and remnants infer grains that are almost exclusively large ($a
  \gtrsim 0.1\;\mu$m)
  \citep{wesson15a,priestley20a,niculescu-duvaz22a}.  Small grains
  from AGB stars are instead destroyed by periodic pulsation-driven
  shocks \citep{winters97a,yasuda12a}.  We therefore initialize newly
  formed dust with lognormal grain size distributions following
  \citet{asano13b}:
\begin{equation}
  \frac{\partial n}{\partial a} = \frac{C}{a^p} \exp\left[ -\frac{\ln^2(a/a_0)}{2\sigma^2}\right]
  \label{equation:initial_gsd}
\end{equation}
where $a$ is the grain radius, $a_0$ is the characteristic (peak)
grain size, $p$ and $\sigma$ are dimensionless shape parameters
controlling the power-law slope and lognormal width, and $C$ is a
normalization constant. In practice, we precompute the shape of
$\partial n / \partial a$, normalize it to unit integrated dust mass
following \citet{narayanan23a}, and then scale by the total dust mass
spawned from the star.  Specifically, $a_0 = 0.1\;\mu$m, and the shape
parameters $(p, \sigma)$ are $(4, 0.47)$ for AGB stars and $(0, 0.6)$
for Type II supernovae \citep{nozawa07a,asano13b}.  We discretize the
grain size distribution into $16$ logarithmic bins spanning $4\times
10^{-4} \leq (a/\mu \mathrm{m}) \leq 1$.  Because PAHs in our
  model are defined as carbonaceous grains with $a < 13$~\AA, and the
  initial size distributions above place a negligible mass fraction at
  these sizes, stellar sources inject effectively no PAHs.  Any PAH
  mass in our model therefore implicitly forms {\it in situ} within
  the ISM (\S~\ref{section:lifecycle}).

If a star particle of mass $M_*$ produces a dust mass $\Delta M_d$
within a timestep $\delta t$, we spawn a new dust particle with
probability:
\begin{equation}
  p_d = \frac{M_*}{M_d}\left[1 - \exp\left(-\frac{\Delta M_d}{M_*}\right)\right]
  \label{equation:spawn_prob}
\end{equation}
Here, $p_d$ is the probability of spawning a new dust super-particle during this timestep.   We split dust particles if their mass exceeds 10 times the target gas mass resolution, and merge them if they fall below 0.1 times this mass \citep{li21a}.

Once spawned, dust particles feel the full gravitational field as
  well as an aerodynamic drag force against the local gas, computed
  from kernel-weighted gas properties over neighboring cells.  We
  adopt Epstein drag (see \S~\ref{section:dust_velocities}) with a correction for
  supersonic dust--gas relative velocities, yielding a stopping
  timescale integrated over the particle's grain size distribution,
  and integrate the drag equation with the semi-implicit exact
  integrator of \citet{lorenaguilar15a}, which remains stable even
  when the stopping time is much shorter than the timestep.  Dust
  timesteps are limited by the minimum hydro timestep of neighboring
  gas cells, a Courant-like criterion on the dust smoothing length and
  effective signal speed, which ensures that dust timesteps are always
  shorter than those of the surrounding gas.

\subsubsection{Dust Growth and Destruction}
\label{section:dust_growth_destruction}

Once in the ISM, dust grains evolve through several concurrent
processes.

{\it Grain growth via metal accretion:} Dust grains grow by accreting
gas-phase metals.  The growth rate follows:
\begin{equation}
  \left(\frac{da}{dt}\right)_{\rm grow} = \frac{a}{\tau_{\rm accr}}
  \label{equation:growth}
\end{equation}
where the accretion timescale is:
\begin{equation}
\begin{split}
  \tau_{\rm accr} = \tau_{\rm ref} & \left(\frac{a}{0.1\;\mu {\rm m}}\right) \left(\frac{1000\;{\rm cm}^{-3} \cdot m_{\rm H} \cdot Z_\odot}{\rho_Z}\right) \\
  & \times \left(\frac{10\;{\rm K}}{T_g}\right)^{1/2} \left(\frac{0.3}{S}\right)
\end{split}
  \label{equation:tau_accr}
\end{equation}
here, $a$ is the grain radius, $m_{\rm H}$ is the mass of hydrogen, $\rho_{\rm Z}$ is the metal density, $T_{\rm g}$ is the gas temperature, $S$ is the sticking coefficient, and $\tau_{\rm ref}$ is a reference timescale.  We assume reference timescales $\tau_{\rm ref} = 0.224$ Gyr for silicates
and $\tau_{\rm ref} = 0.175$ Gyr for carbonaceous grains \citep{hirashita00a}, and a
temperature-dependent sticking coefficient $S$ from
\citet{zhukovska16a} that reduces growth efficiency in warm gas.

{\it Thermal sputtering:} In hot gas, collisions with thermal ions
erode grain surfaces, at a rate:
\begin{equation}
  \left(\frac{da}{dt}\right)_{\rm sp} = -\frac{a}{\tau_{\rm sp}}
  \label{equation:sputtering}
\end{equation}
where the sputtering timescale follows the analytic approximation of
\citet{tsai95a}:
\begin{equation}
\tau_{\rm sp} = 0.17 \;{\rm Gyr}\;\left(\frac{a}{a_{\rm ref}}\right)\left(\frac{10^{-27}\;{\rm g\;cm}^{-3}}{\rho_g}\right)\left[\left(\frac{T_0}{T}\right)^\omega + 1\right]
\end{equation}
with $a_{\rm ref} = 0.1\;\mu$m a reference grain radius, $\rho_g$ the local gas density, $T$ the local gas temperature, $\omega = 2.5$ and $T_0 = 2\times10^6$ K, where $T_0$ sets the temperature scale above which thermal sputtering becomes efficient and $\omega$ controls the steepness of the sputtering rate with gas temperature.

{\it Supernova shock destruction:} Dust grains swept up by supernova
blast waves undergo partial or complete destruction.  We model the
grain-size-dependent destruction efficiency following
\citet{nozawa06a} and \citet{asano13b}, where we determine the mass swept by
shocks from each SN event following the prescriptions
of \citet{yamasawa11a}.

Sputtering and shock destruction reduce the mass of the dust
  super-particle; the lost mass is returned to the surrounding gas as
  gas-phase metals, distributed element-by-element over the
  neighboring cells with volume-weighted kernel weights, ensuring mass
  conservation.  Grain growth via accretion is the reverse operation,
  drawing from the gas-phase metal content of neighboring cells and
  limited by the metals available in each cell.  Dust particles whose
  mass reaches zero are removed from the simulation.

{\it Grain-grain collisions:} The interaction of dust grains with one
another drives both shattering and coagulation, following the
framework of \citet{mckinnon18a} and \citet{li21a}.  The rate of mass
change in grain size bin $k$ due to collisions is:
\begin{eqnarray}
  \label{equation:shattering}
  \frac{dM_k}{dt} &=& -\frac{\pi \rho_d}{M_d} \Biggl(\sum_{i} v_{\rm rel}(a_i, a_k)\; m_i \; I^{i,k} \nonumber \\
  & & - \frac{1}{2}\sum_{i}\sum_{j} v_{\rm rel}(a_i, a_j)\; m_{\rm col}^{i,j}(k) \; I^{i,j}\Biggr)
\end{eqnarray}
where $\rho_d$ is the local dust density, $M_d$ is the total dust mass
of the dust particle, $v_{\rm rel}(a_i, a_k)$ is the relative velocity
between grains of sizes $a_i$ and $a_k$, $m_i$ is the average grain
mass in bin $i$, and $I^{i,k} = \int (a_i + a_k)^2 n_i n_k \; da_i
\; da_k$ is the collision cross-section integral (with $n_i$ and $n_k$ the grain number densities in bins $i$ and $k$).   $m_{\rm col}^{i,j}(k)$ is the mass that collisions between grains in bins $i$ and $j$ deposit into bin $k$, either as shattered fragments or as a coagulated product.  The first term
represents the mass lost from bin $k$ due to collisions with all other
sizes, while the second term represents the mass gained in bin $k$
from the fragments or merged products of collisions between other
bins.

Whether a given collision results in shattering or coagulation depends
on the relative velocity compared to material-dependent threshold
velocities.  For shattering, a collision occurs when $v_{\rm rel} >
v_{\rm shatter}$, where $v_{\rm shatter} = 2.7$ km s$^{-1}$ for silicates
and $v_{\rm shatter} = 1.2$ km s$^{-1}$ for carbonaceous grains
\citep[][though see \citet{esmerian26a} for potential updates to these values.]{jones96a}.  
For coagulation, grains stick when $v_{\rm rel} < v_{\rm coag}(a_i,
a_k)$, where the coagulation threshold is a size-dependent quantity
that depends on the surface energy, elastic modulus, and internal
grain density of the dust species \citep[Equation 8
  of][]{hirashita09a}.  Smaller grains have higher coagulation
thresholds (and are thus more easily coagulated), while larger grains
require lower relative velocities to stick.  We deposit the combined
mass of two coagulated grains into the appropriate bin for the merged
grain size.

The local turbulent cascade sets the relative velocity between two grains in a given size bin; we assume the cascade is driven at the scale of
the local Jeans length, $L_{\rm J}$.  Grains acquire inertial drift
velocities as they decouple from turbulent eddies; the extent of this
decoupling is quantified by the Stokes number, $St = \tau_s /
\tau_{\rm eddy}$, where $\tau_s$ is the grain stopping time in the
Epstein drag regime. 
We provide a full derivation of the dependency of the collision velocity
on local ISM physical conditions, and its consequences for shattering
in \S~\ref{section:dust_velocities}, where we also characterize the
dominant physical parameters that control the turbulent velocity, $v_b$.

In the remainder of this paper, we assume that PAHs are
carbonaceous-dominated dust grains with fewer than $10^3$ carbon
atoms, corresponding to grain sizes $a < 13$ \AA \citep{hensley22a}.  We note that we do not include a model for aliphatic and aromatic grains separately.   The processes that drive the conversion between aliphatics and aromatics are fairly uncertain, and can include dependencies on the interstellar radiation field, local hydrogen density, and removal of aliphatic side groups from aromatics.  We defer an exploration of the dependency of our model on these processes to a future work.

Finally, we emphasize that this treatment is an effective approximation:
  in reality, PAHs are likely molecules \citep[e.g.][]{tielens08a},
  not solid grains.  By modeling them as the smallest tail of the
  carbonaceous grain size distribution, we implicitly assume that PAHs
  participate in the same collisional network (shattering and
  coagulation) as the broader grain population, and that their
  dynamics are fully coupled to it.  Beyond this, we do not model
  processes that would otherwise operate uniquely on molecules, such
  as photo-ionization, photo-dissociation, chemistry, or
  molecule-molecule reactions.  The molecular nature of PAHs does,
  however, enter our framework in the computation of their emission,
  in which each PAH is treated as an individual molecule
  stochastically excited by single-photon absorptions
  (\S~\ref{section:powderday}).

\subsection{PAH Heating and Emission Calculations}
\label{section:powderday}

To compute the emergent mid-IR spectra of our model galaxies, we
post-process each simulation snapshot through the {\sc powderday} dust
radiative transfer package \citep{narayanan21a}.  {\sc powderday}
couples the {\sc yt}-based grid generation \citep{turk11a}, stellar
population synthesis from {\sc fsps}
\citep{conroy09a,conroy10a,conroy10b} using MIST stellar isochrones
\citep{choi16a}, and Monte Carlo radiative transfer via {\sc hyperion}
\citep{robitaille11a}.  We assign each star particle a spectral energy
distribution based on its age and metallicity: each star particle
  is treated as a simple stellar population with a \citet{kroupa02a}
  initial mass function, with its metallicity matched to the nearest
  point on the MIST grid.  The
  spectra combine the empirical MILES library
  \citep{sanchezblazquez06a} in the optical with the BaSeL 3.1
  theoretical atmospheres \citep{lejeune97a,lejeune98a,westera02a} at
  all other wavelengths.  We consider wavelengths longward of the Lyman Limit in the radiative transfer calculations.  We
propagate photon packets isotropically through a Voronoi mesh
constructed around the dust particles.  We compute the extinction in each cell
self-consistently from the local grain size distribution and
composition, using extinction efficiencies from \citet{draine84a}
(silicates) and \citet{laor93a} (carbonaceous grains).  Once a
  photon enters a cell, the geometry is uniform, so the dust is
  treated as a uniform screen, with no subresolution clumping or
  geometry assumed.  Photon packets scatter or absorb based on the
local albedo, and we iterate the radiative transfer until the energy
absorbed by each cell converges to better than 1\%.  We describe
  the construction of the Voronoi grid from the dust particles, the
  photon transport and optical depth computation through it in
  Appendix~\ref{appendix:rt_details}.  A key output of
this procedure is the mean interstellar radiation field (ISRF) in each
cell (for wavelengths $\lambda > 912$ \AA), which we use to compute
the PAH emission.

We parameterize the strength of the local radiation field via the dimensionless intensity parameter $U$, defined as the rate of energy absorption by dust grains normalized to that in the solar neighborhood \citep{mathis83a,draine21a}:
\begin{equation}
 \label{equation:logu}
  U \equiv \frac{\int c \, u_\lambda \, C_{\rm abs}(\lambda) \, d\lambda}{h_{\rm ref}}
\end{equation}
where $c \, u_\lambda \, C_{\rm abs}(\lambda)$ is the rate of energy absorption per unit wavelength for a grain with absorption cross-section $C_{\rm abs}(\lambda)$ in a radiation field with energy density $u_\lambda$, and $h_{\rm ref} = 1.958 \times 10^{-12}$ erg s$^{-1}$ is the corresponding heating rate in the modified \citet{mathis83a} (mMMP) interstellar radiation field.  Thus $U = 1$ corresponds to solar neighborhood conditions, while star-forming regions typically have $U \gg 1$.

\subsubsection{PAH Emission via the Single Photon Approximation}
\label{section:helena}

\begin{figure*}
  \centering
  \includegraphics[width=\textwidth]{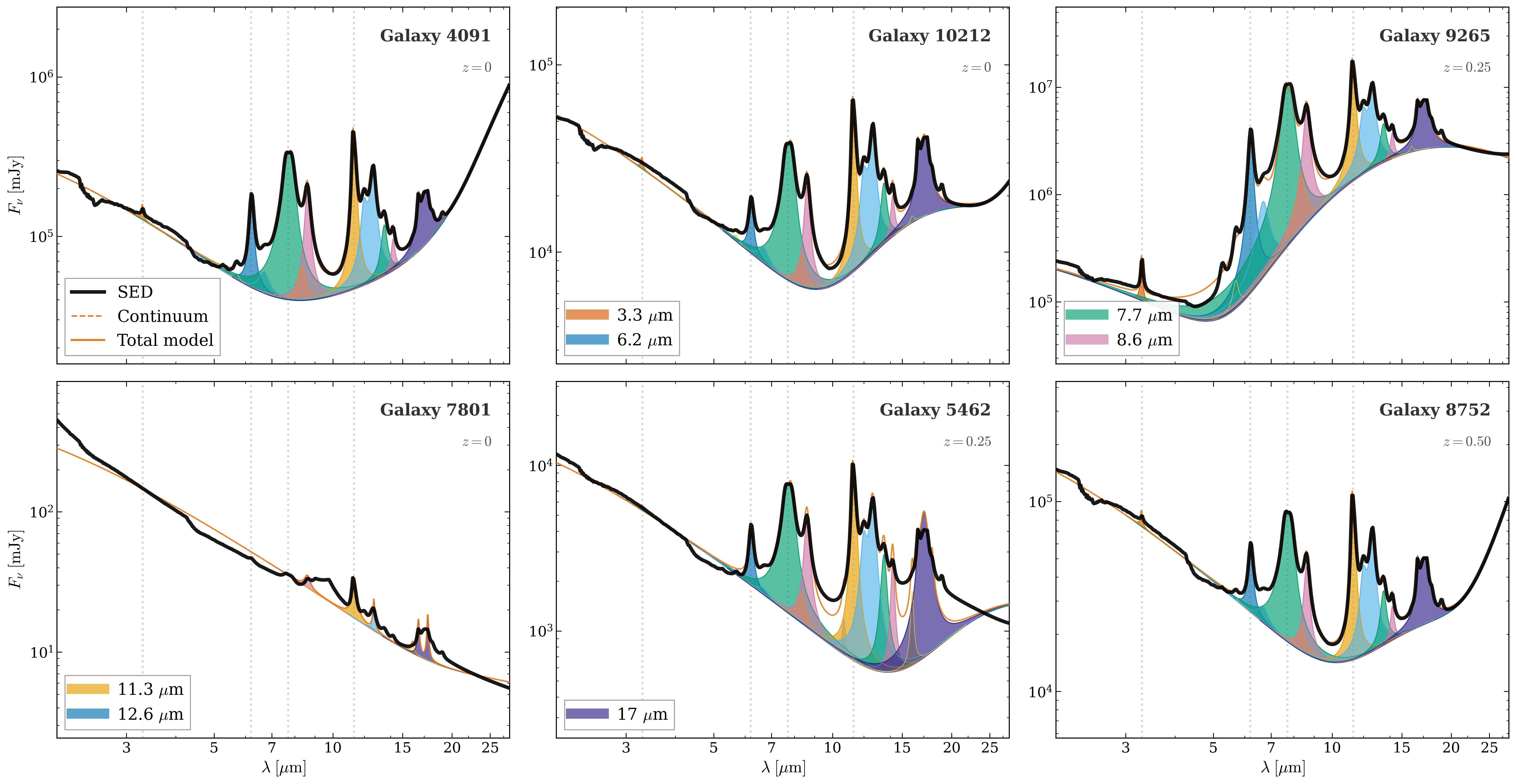}
  \caption{{\bf Mid-infrared spectral decomposition of six
      representative simulated galaxies at $z \leq 0.5$ using {\sc
        pahfit}.} Each panel shows the rest-frame mid-IR SED (grey)
    along with the {\sc pahfit} best-fit decomposition.  The solid orange
    curve shows the total model (continuum plus all PAH features).
    Individual PAH feature families are shown as color-filled regions
    stacked on the continuum, with the major complexes at $3.3$,
    $6.2$, $7.7$, $8.6$, $11.3$, $12.7$, and $17\;\mu$m indicated in
    the legend.  Vertical dotted lines mark the $3.3$, $6.2$, $7.7$,
    and $11.3\;\mu$m feature positions for reference.
    \label{figure:sed_decomposition}}
\end{figure*}

With the ISRF in hand, we compute the PAH emission spectrum in each
cell using the single photon approximation (SPA) model of
\citet{richie25a}, as implemented in {\sc powderday}.  PAH molecules
are sufficiently small that they do not typically reach thermal equilibrium with
the ambient radiation field \citep{draine11a,li20b}.  Instead, they  
operate in a stochastic heating regime.  Here, the absorption of a single UV
photon vibrationally excites the molecule to a peak temperature
$T_{\rm peak}$, after which it cools radiatively via emission in the
classical PAH bands \citep{leger84a,allamandola85a}. We summarize the key elements
of the \citet{richie25a} model pertinent to our simulations here.  It
is important to note that this is a significant update from the
modeling procedure outlined in \citet{narayanan23a}, which modeled the
ISRF as a superposition of the pre-computed stellar SEDs by
\citet{draine21a}, and tied the emergent spectra to linear
combinations of the \citet{draine21a} pre-computed SEDs.

The core quantity in the SPA is the {\it basis spectrum}
$\tilde{p}_{\lambda_{\rm em}}(\lambda_{\rm abs})$: the time-averaged power emitted
per unit wavelength during a single complete cooling event.  When a
PAH of size $a$ absorbs a photon of wavelength $\lambda_{\rm abs}$, it
receives energy $E_{\rm abs} = hc/\lambda_{\rm abs}$ and is heated to
a peak temperature $T_{\rm peak}$.  The grain then cools radiatively
according to:
\begin{equation}
  \frac{dE}{dt} = -\int_{\lambda_{\rm abs}}^{\infty} 4\pi \, B_\lambda\bigl(T(t)\bigr) \, C_{\rm abs}(\lambda) \, d\lambda
  \label{equation:spa_cooling}
\end{equation}
where $B_\lambda$ is the Planck function and $C_{\rm abs}(\lambda)$ is
the grain absorption cross-section \citep[from][]{draine21a}.
\citet{richie25a} compute the internal energy $E(T)$ by summing over
the vibrational modes of the PAH molecule, treated as quantum harmonic
oscillators.  This cooling equation is integrated from $T_{\rm peak}$
to $T=0~{\rm K}$ using the thermal continuous approximation at high
energies \citep{draine01a} with adaptive energy binning ($dE/E=3\%$
per step), and pre-defined bins based on PAH vibrational mode energies
for the ten lowest bins. The resulting temperature history $T(t)$ is
used to construct the basis spectrum.  The basis spectrum encodes
which mid-IR features receive power for a given absorbed photon
energy: harder photons drive the grain to higher $T_{\rm peak}$,
shifting emission toward shorter wavelengths and enhancing the
$3.3\;\mu$m feature relative to longer wavelength features.  We
pre-compute these basis spectra for the specific dust grain sizes
modeled here, and absorbed photon wavelengths from $912$ \AA\ to
$10.1\;\mu$m across $474$ logarithmic bins \citep{richie25a}.  It is
important to note in the current implementation of \citet{richie25a},
large PAHs may have their temperature distributions incorrectly
estimated at log($U$)$ \ga 2$.  This said, these conditions are
uncommon in our simulations given our inability to resolve individual
HII regions, and we show later in this paper
(\S~\ref{section:pah_sfr}) that our dust-mass weighted average
log$\left(U\right)$ values do not typically exceed this threshold.

We then construct the full emission spectrum for a grain of size $a$ in a radiation
field with energy density $u_\lambda$ by weighting
the basis spectrum by the rate of photon absorption in each wavelength
channel.  Since the emission
from each absorbed photon is independent, we can construct the full PAH spectrum
as a simple linear combination of pre-computed basis
spectra.

Neutral and ionized PAHs have distinct absorption cross-sections
$C_{\rm abs}(\lambda)$, and therefore distinct basis spectra.  We
compute spectra for both ionization states independently.  The grain
ionization fraction as a function of size follows the analytic
relation of \citet{draine21a} and \citet{hensley22a}:
\begin{equation}
  f_{\rm ion}(a) = 1 - \frac{1}{1+a/(10\;\text{\AA})}
  \label{equation:fion}
\end{equation}
such that the smallest grains are predominantly neutral while larger
PAHs are increasingly ionized.  The total PAH emission per cell is
then:
\begin{equation}
  p_\lambda^{\rm tot} = \sum_a \left[ f_{\rm ion}(a) \, p_\lambda^+(a) + \bigl(1 - f_{\rm ion}(a)\bigr) \, p_\lambda^0(a) \right] n(a)
  \label{equation:pah_total}
\end{equation}
where $p_\lambda^+$ and $p_\lambda^0$ are the ionized and neutral
emission spectra, and $n(a)$ is the grain number density from the
simulation's grain size distribution.

Equation~\ref{equation:fion} is an approximation, included to
  accommodate the fact that we cannot directly model PAH charge states
  in our simulations.  The ionization state of a PAH is set by a balance
  between photoionization by the local FUV field, and recombination
  with free electrons.  Equation~\ref{equation:fion} encodes this
  balance averaged over a standard
  mixture of diffuse ISM phases \citep{draine21a,hensley22a}.
  Accurately modeling the free electron density requires a coupling
  with ionizing radiative transfer, which we defer to a future study.  The practical consequence of the approximation embedded in Equation~\ref{equation:fion} is to impact the PAH band ratios \citep{draine21a}, though the total PAH luminosities (the subject of this paper) are insensitive to this approximation.

In practice, for each cell in the {\sc powderday} Voronoi mesh, we convert the
local ISRF from the Monte Carlo radiative transfer to an
energy density $u_\lambda$, 
and the SPA calculation returns the total PAH luminosity per unit
wavelength.  We then add the resulting PAH emission spectra as source
terms in the radiative transfer, and iterate the dust radiative
transfer a second time to capture any attenuation of the PAH features
in extremely dense regions.  Following \citet{richie25a}, the
  emission profiles of the PAHs are Drude profiles \citep{draine21a}.
  We obtain the final emergent SED for each galaxy using the peeling-off technique as implemented in {\sc hyperion} \citep{robitaille11a}.    We
    compute the SED along a single viewing direction per snapshot.  Because galaxy orientations in our zoom simulations are     random with respect to this direction, our
  sample statistics effectively sample random inclinations.   This
  process allows us to capture both the PAH mid-IR emission, as well
  as the broadband UV-millimeter wave SED, including thermal
  far-infrared dust emission.

\subsubsection{Fitting Procedures and Extraction of PAH Feature Strengths}
\label{section:pahfit}
To extract PAH feature luminosities from our simulated mid-IR spectra,
we employ {\sc pahfit} \citep{smith07a}, a spectral decomposition tool
designed to separate the blended components of mid-infrared emission.
We use {\sc pahfit} in order to compare our results in as
  consistent of a manner as possible to observations, as well as to
  derive the individual fluxes of separated PAH features.  {\sc
  pahfit} models the observed spectrum as a superposition of starlight
(represented as a $T = 5000$ K blackbody), thermal dust continuum
emission (a sum of modified blackbodies at fixed temperatures spanning
$35$--$300$ K), atomic and molecular gas emission lines, and dust
emission features.  The features/complexes modeled by {\sc
    pahfit} are the exact same as those modeled in our {\sc powderday}
  simulations, to ensure no spurious modeling of emission features.
We apply attenuation by silicate absorption at $9.7$ and $18\;\mu$m
using the mixed-geometry extinction model of \citet{smith07a}.

We parameterize the PAH emission features as Drude profiles \citep{draine07a,smith07a}:
\begin{equation}
  I_\nu(\lambda) = \frac{b\,\gamma^2}{\left(\lambda/\lambda_0 - \lambda_0/\lambda\right)^2 + \gamma^2}
  \label{equation:drude}
\end{equation}
where $I_\nu(\lambda)$ is the specific intensity of an individual
  PAH feature as a function of wavelength $\lambda$, $b$ is the peak
intensity, $\lambda_0$ is the central wavelength, and $\gamma$ is the
width.  {\sc pahfit} fits the major PAH complexes at $3.3$, $6.2$,
$7.7$, $8.6$, $11.3$, $12.7$, and $17\;\mu$m, as well as weaker
features at $5.3$, $5.7$, $6.7$, $10.7$, $12.0$, $14.0$, and
$33.1\;\mu$m.  Several of these are blended complexes composed of
multiple sub-features: the $7.7\;\mu$m complex is decomposed into
three Drude components centered at $7.42$, $7.60$, and $7.85\;\mu$m,
the $11.3\;\mu$m complex into two components at $11.23$ and
$11.33\;\mu$m, and the $12.7\;\mu$m complex into two components at
$12.62$ and $12.69\;\mu$m.  We hold the central wavelengths and widths
of the Drude profiles fixed during fitting, allowing only the
amplitudes to vary, following typical practice for observational
  galaxy mid-IR spectral decomposition.

We define the total PAH luminosity $L_{\rm PAH}$ as the sum of the
integrated powers of all PAH Drude features returned by {\sc pahfit}.
We compute individual feature luminosities (e.g., $L_{3.3}$, $L_{7.7}$,
$L_{11.3}$) by summing the powers of the Drude
sub-components belonging to each complex.

In Figure~\ref{figure:sed_decomposition}, we show representative {\sc
  pahfit} decompositions of six simulated galaxy mid-IR spectra at $z
\leq 0.5$.  The individual PAH feature families are shown as
color-filled regions atop the dust continuum baseline (dashed line),
and the total model (solid line) is the sum of the continuum and all
PAH features.  The decompositions cleanly separate the PAH emission
from the underlying continuum across a diverse range of galaxies,  and demonstrate that our simulated spectra
produce realistic PAH feature profiles that are well-described by the
Drude profile parameterization.

 \section{Part I: The Physical Conditions Driving The Formation of PAHs}
In this section, we examine the formation and evolution of PAHs
  in our simulated galaxies from a physical perspective (i.e., without
  considering their luminous properties). In particular, we focus on
  the ISM properties and processes responsible for PAH production and
  their overall impact on PAH abundances as galaxies evolve from
  $z=6\rightarrow 0$.
 
\label{section:lifecycle}
\subsection{The Physical Conditions driving Dust Velocities}
\label{section:dust_velocities}
Because dust is large upon formation in our model
(\S~\ref{section:dust_production}), PAHs must form {\it in situ} via
grain-grain shattering.  The shattering rate depends on the relative
velocity between grains, which is itself set by local gas conditions.
We now derive this relationship explicitly, as it will inform how and
when shattering occurs in galaxies.

We assume that grain motions within individual dust super-particles
are driven by an unresolved turbulent cascade, and that the outer scale of the unresolved energy cascade is the local Jeans length, $L_{\rm J}$. We further assume that the velocity dispersion scales with
the size scale of turbulence according to \citet{larson81a}'s scaling relations:
\begin{equation}
  v_{\rm turb} \propto L_{\rm J}^{1/2}
\end{equation}
Where $v_{\rm turb}$ is the turbulent velocity and $L_{\rm J}$ is the Jeans length.  The Jeans length is given by:
\begin{equation}
  \label{equation:jeans}
  L_{\rm J} \sim  \frac{c_s}{\sqrt{G \rho}} \sim  T^{1/2} \rho^{-1/2}
\end{equation}
\ifshowcode
\begin{verbatim}
/* grain_sizes.c: Line 1009 */
double Lmax = 0.5 * sqrt(M_PI * cs * cs / (GRAVITY * rho_local_cgs)) / PARSEC;
\end{verbatim}
\fi
where $c_s$ is the local isothermal sound speed, $G$ is the gravitational constant, $\rho$ is the local gas density, and $T$ is the local gas temperature.
Then:
\begin{equation}
v_{\rm turb} \sim  L_{\rm J}^{1/2} \sim  \left(T^{1/2} \rho^{-1/2}\right)^{1/2} \sim  T^{1/4} \rho^{-1/4}
\end{equation}
\ifshowcode
\begin{verbatim}
/* grain_sizes.c: Line 1011 */
*v_turb = 7e8/2.355 * pow(Lmax,0.5);
\end{verbatim}
\fi
where we have used the relation $c_s \sim T^{1/2}$ for the isothermal sound speed.

Let us now consider the motion of a dust grain of radius $a$ and
internal density $\rho_{\rm grain}$ embedded in this turbulent ISM.  Here, turbulent eddies drive an acceleration on the dust grains, which we model as centrifugal acceleration:
\begin{equation}
    a_{\rm turb} \sim \frac{v_{\rm turb}^2}{L_{\rm J}}.
    \label{equation:accel}
\end{equation}
In the Epstein drag regime (where grains are smaller than the gas mean
free path), the stopping time of a grain is $\tau_s \sim \rho_{\rm
  grain} a / (\rho c_s)$ \citep{draine79b,mckinnon18a}.  The grain
achieves an equilibrium relative velocity, $v_{\rm b}$, when drag
balances the turbulent acceleration
\citep{yan04a}:
\begin{equation}
v_{\rm b} \approx \tau_s \, a_{\rm turb}.
\end{equation}
Substituting $\tau_s$ and Equation~\ref{equation:accel} into this expression:
\begin{equation}
  \label{equation:vb1}
    v_b \sim \left( \frac{\rho_{\rm grain} a}{\rho c_s} \right) \left( \frac{v_{\rm turb}^2}{L_{\rm J}} \right).
\end{equation}


We can rearrange Equation~\ref{equation:vb1} in terms of the Mach number, the dimensionless ratio of the turbulent velocity to the local sound speed:
\begin{eqnarray}
  \mathcal{M} &=& \frac{v_{\rm turb}}{c_s} \nonumber \\
    &\sim& \frac{T^{1/4} \rho^{-1/4}}{T^{1/2}} \nonumber \\
    &\sim& T^{-1/4} \rho^{-1/4}.
    \label{equation:mach}
\end{eqnarray}
\ifshowcode
\begin{verbatim}
/* grain_sizes.c: Line 1012 */
mach = *v_turb / (cs * cm_to_um);
\end{verbatim}
\fi

This results in:
\begin{eqnarray}
    v_b &\sim& \frac{\rho_{\rm grain} a}{\rho L_{\rm J}} \left( \frac{v_{\rm turb}^2}{c_s} \right) \nonumber \\
    &\sim& \frac{\rho_{\rm grain} a}{\rho L_{\rm J}} \mathcal{M}^2 c_s.
    \label{equation:drift_mach_explicit}
\end{eqnarray}
This intermediate step recovers the key result from \citet{yan04a}:
the grain velocity scales with the square of the Mach number, $v_b
\sim \mathcal{M}^2$.

Finally, to find the final dependence on gas density and temperature,
we substitute the scaling relation for $L_{\rm J}$
(Equation \ref{equation:jeans})  back into
Equation \ref{equation:drift_mach_explicit}:
\begin{eqnarray}
    v_b &\sim& \frac{\rho_{\rm grain} a}{\rho (c_s \rho^{-1/2})} \mathcal{M}^2 c_s \nonumber \\
    &\sim& \frac{\rho_{\rm grain} a}{\rho^{1/2}} \mathcal{M}^2.
\end{eqnarray}
Next, we substitute $\mathcal{M}^2 \sim (T^{-1/4} \rho^{-1/4})^2 = T^{-1/2} \rho^{-1/2}$:
\begin{eqnarray}
  v_b &\sim& \frac{\rho_{\rm grain} a}{\rho^{1/2}} \left( T^{-1/2} \rho^{-1/2} \right) \nonumber
\end{eqnarray}
Combining terms yields the final scaling relation:
\begin{equation}
  \label{equation:vb_rhot}
    v_b(a, \rho, T) \sim a \, \rho_{\rm grain} \, T^{-1/2} \, \rho^{-1}.
\end{equation}

\ifshowcode
\begin{verbatim}
/* grain_sizes.c: Line 1017*/
v_b = 9e7 * pow(mach,2) * (a/0.1) * pow(rho_H_ref/rho_H_local,0.5) 
      * (DARR(All.GrainDensity,s)/3.5);
\end{verbatim}
\fi

\begin{figure*}
\centering
  \includegraphics[scale=0.4]{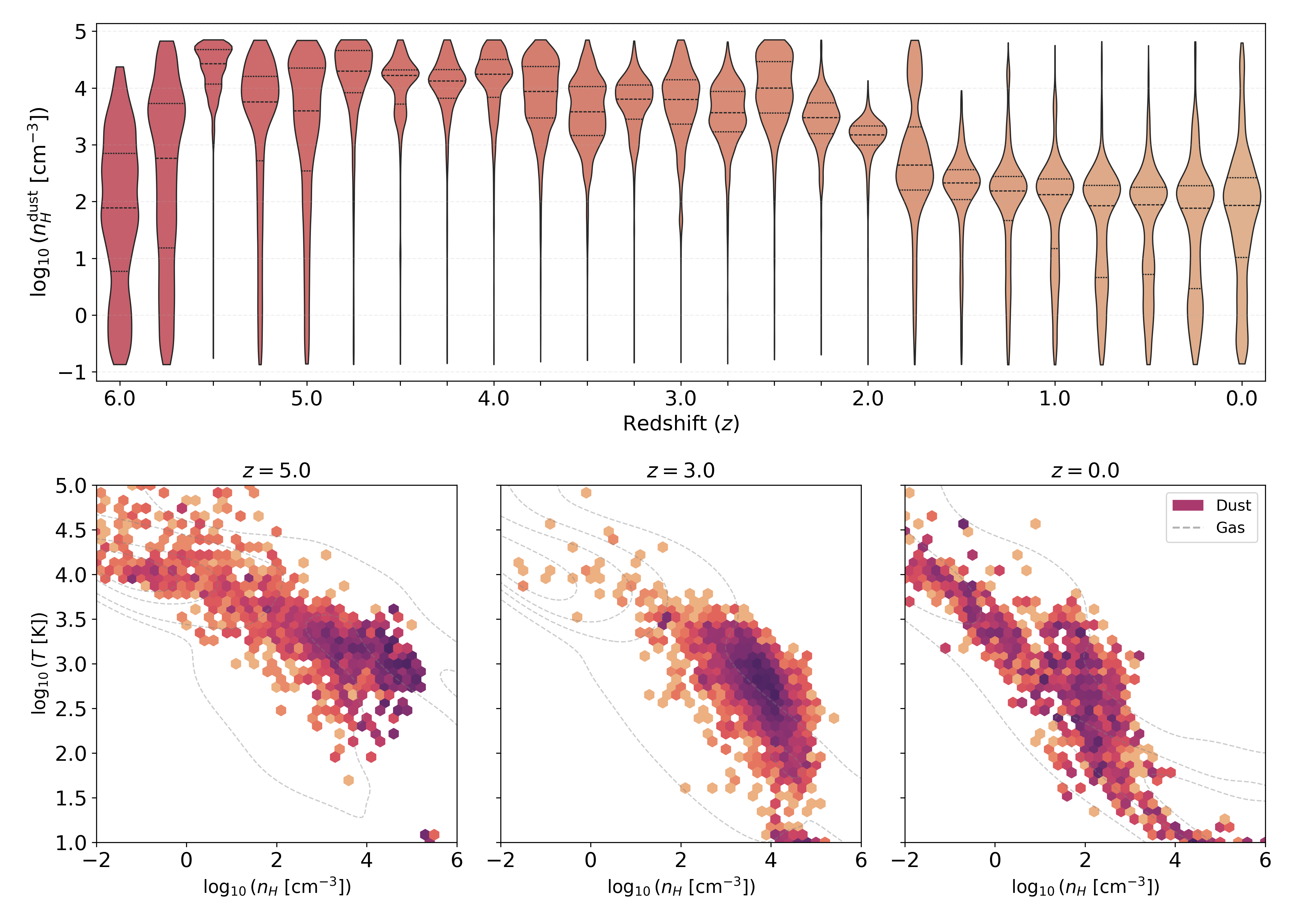}
  \caption{{\bf The evolution of the ISM physical conditions in the
      vicinity of dust over cosmic time.}  {\it Top:} Violin plot
    showing the distribution of gas densities for gas particles that
    neighbor dust particles in our model galaxy sample.  Broadly,
    toward late times ($z\la 4$ in this case), the average gas
    densities decrease with time.  The horizontal lines show the median and interquartile ranges. {\it Bottom:} $\rho_{\rm
      gas}-T_{\rm gas}$ plot for gas particles surrounding dust
    particles (colored), and all gas particles (dashed contours) at $z=5,3,0$ for our model galaxy sample.  The colors are a heat map representing the density of points.
    \label{figure:dust_density_temp_z}}
\end{figure*}

The inverse dependence on density is the strongest effect. In dense
gas, the stopping time is short ($\tau_s \propto \rho^{-1}$) because
the frequent collisions with gas atoms efficiently transfer
momentum. This creates a tight-coupling regime where grains are
well-coupled with the gas flow, resulting in
negligible relative velocities.   In diffuse
gas, the coupling is weak, allowing grains to decouple from the gas
eddies and acquire significant inertial velocities. The inverse
dependence on temperature is counter-intuitive: the kinetic energy of
the gas is of course higher at large temperatures, though the velocity
of the dust moving {\it through} that gas does not follow the same
temperature dependence.  This is due to the fact that ({\it i}) In the
Epstein drag regime, there is more drag in warmer gas (due to
increased collisions), resulting in more aligned dust-gas velocities,
and ({\it ii}) in colder gas, $\mathcal{M}$ increases, resulting in higher
turbulent velocities.

\subsection{The Cosmic Evolution of the Physical Conditions Surrounding Dust}
Having computed the relevant physical conditions for driving
shattering in our model, we now turn to calculating the redshift
evolution of the typical densities and temperatures in the gas
surrounding dust particles in our simulations.  In
Figure~\ref{figure:dust_density_temp_z}, we compute the physical
conditions of the nearest $64$ gas particles to every dust particle in
every model galaxy that we simulate, and plot the median and
dispersion of the density  as a function of redshift as
a violin plot.

The average density of the environment that dust resides in rises from
$z \approx 6$ to $z \sim 4$, and then declines toward present
epoch\footnote{We take this opportunity to emphasize that
Figure~\ref{figure:dust_density_temp_z} constitutes a prediction {\it
  only} for the range and distribution of galaxy masses that we have
simulated.  Because these are not large-box cosmological simulations,
they do not constitute a prediction for the inferred physical
properties from a survey of galaxies.}.  The growth in the diffuse to
dense ratio at $z<4$ is driven by a combination of the growth of
galaxy disks with cosmic time, reducing the typical ISM pressure, as
well as a reduction in molecular gas due to star formation (coupled
with the decline in intergalactic gas accretion).  This is a well
established theoretical result that has been seen in a diverse range
of simulation methodologies, including semi-analytic models
\citep[e.g.][]{obreschkow09a,lagos11a}, semi-empirical
\citep{popping15a}, and bona fide hydrodynamic simulations
\citep{dave20a}. This evolution is additionally supported by
observational constraints that demonstrate that while the cosmic
H\,\textsc{i} density evolves weakly \citep[e.g.][]{catinella18a}, the
molecular gas density drops precipitously from $z \sim 2$ to $z=0$
\citep{tacconi13a, saintonge17a, aravena19a}.  At the same time, the
decrease in average SFR in galaxies at redshifts $z \la 2-3$
\citep{madau14a} results in less dust mass being formed in cold, dense
gas at late times; the result is that extant dust has more time to
migrate to warmer, more diffuse conditions.

In the bottom panel of Figure~\ref{figure:dust_density_temp_z}, we
explicitly show the redshift evolution of the physical conditions
surrounding dust in our model galaxies in $\rho-T$ space at $3$
individual redshifts: $z=[5,3,0]$.  Here, the colored hex-bins show
the density of dust particles, while the background contours show the
location of all of the gas in the galaxy in $\rho-T$ space.  The
typical density near dust drops by $\sim 2$ orders of magnitude
between $z=5\rightarrow0$, which will correspondingly increase the
typical relative velocity between dust particles
(cf. Equation~\ref{equation:vb_rhot}).

\begin{figure}
  \includegraphics[scale=0.25]{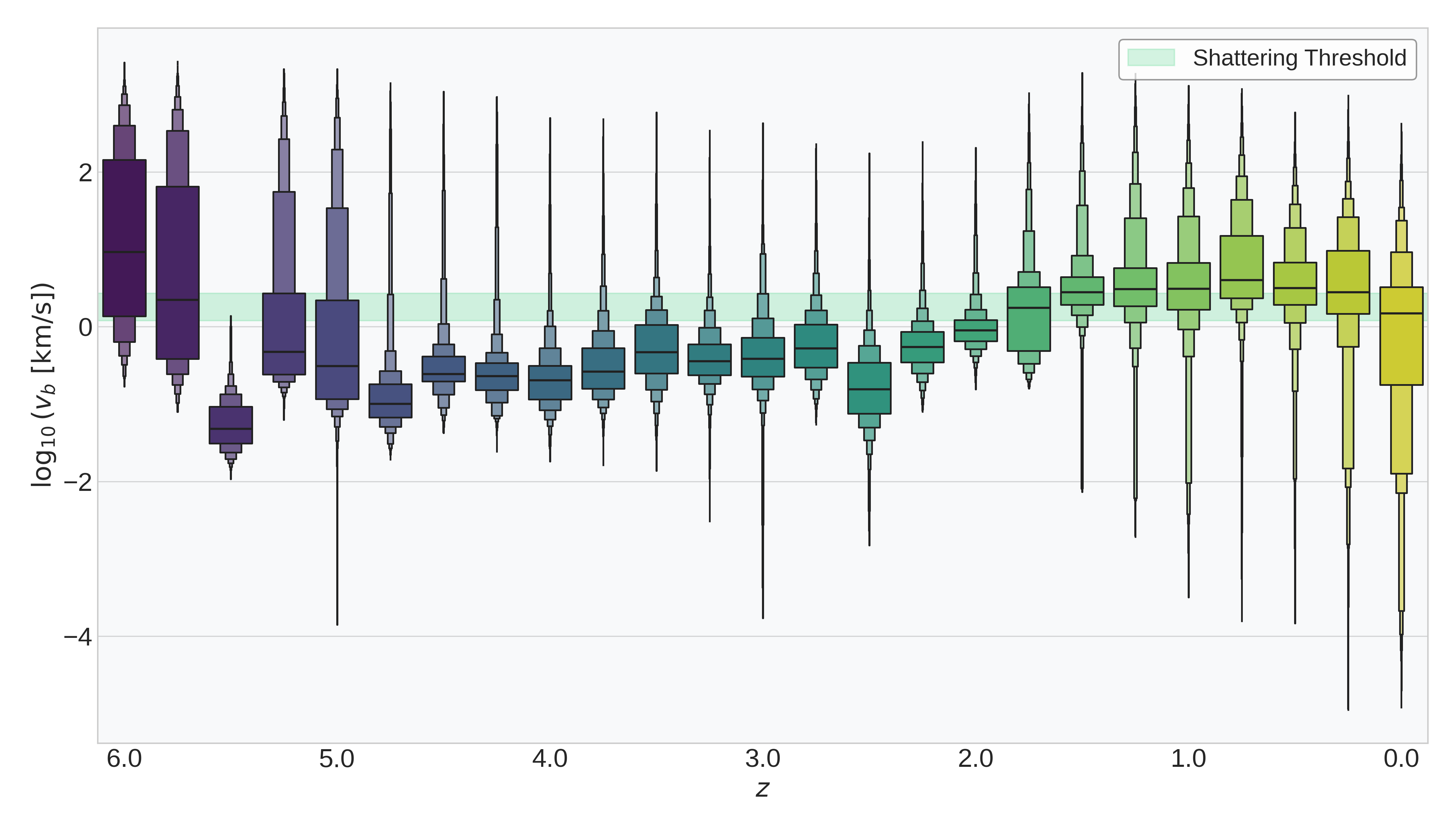}
  \caption{{\bf Distribution of dust grain collision velocities as a
      function of redshift for model galaxy sample.}  The boxenplots
    show the mass-weighted statistics (weighting by grain mass across
    size bins).  The shaded region shows the range of shattering
    threshold velocities that we assume for silicates and carbonaceous grains.  The increase in grain-grain
    collision velocity at $z\la 4$ is a direct consequence of the evolution of ISM
    physical conditions with redshift
    (cf. Figure~\ref{figure:dust_density_temp_z}). \label{figure:vdisp_z}}
\end{figure}

\subsection{The Formation of PAHs via Grain-Grain Shattering}

\begin{figure}
  \includegraphics[scale=0.4]{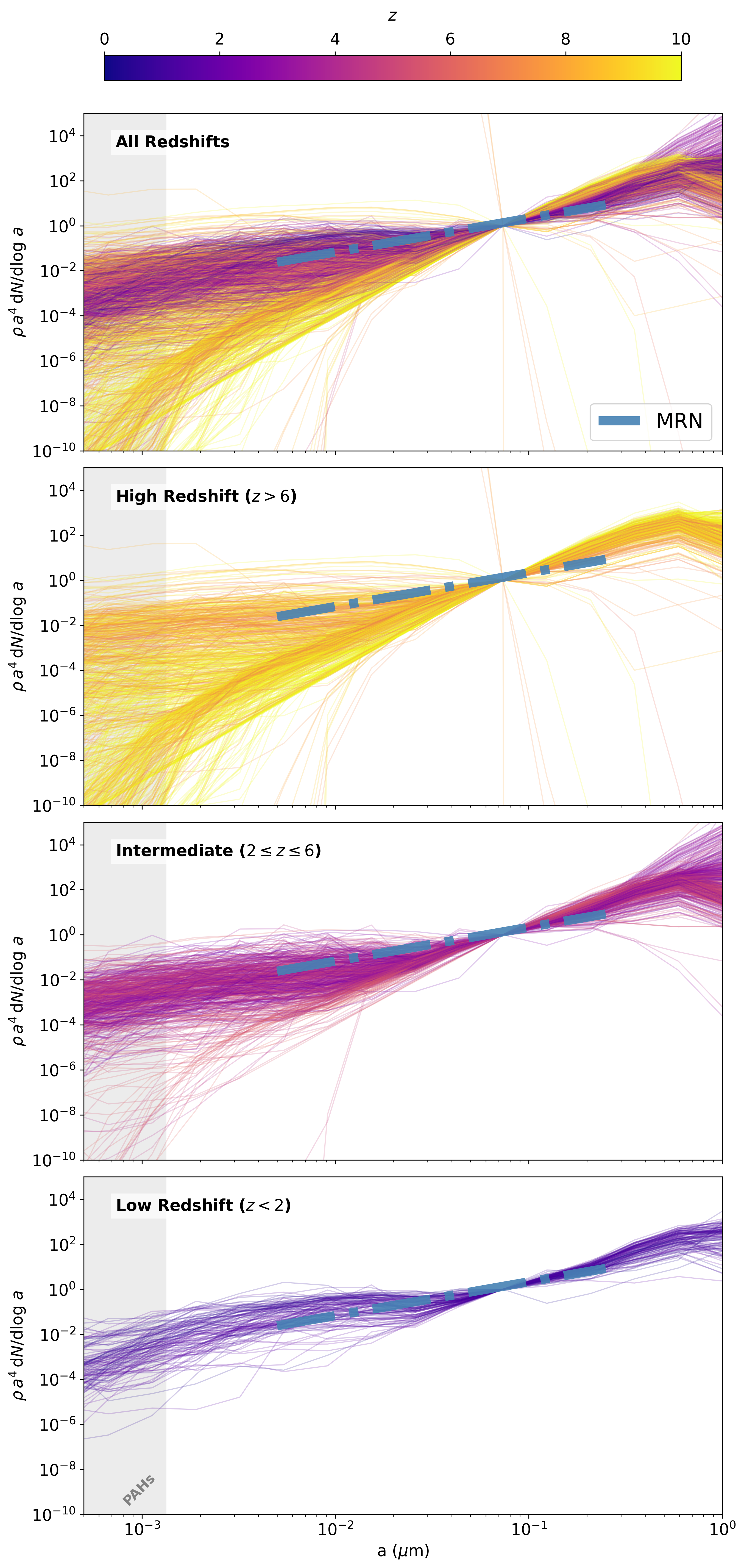}
  \caption{{\bf Evolution of normalized grain size distribution for our
      model sample of galaxies with cosmic time.}  Galaxies are binned
    (bottom to top) at $z<2$, $2 \leq z \leq 6$, and $z>6$, and all
    galaxies are shown up top, color coded by redshift.  For reference, the \citet{mathis77a} grain size distribution is shown in all panels with a thick dash-dot line.  Ultrasmall
    grains that could be considered PAHs are shown in the grey shaded
    region, though we note that the grain size distribution is shown
    for all dust (silicates and carbonaceous grains).  As the typical
    interaction velocities increase with decreasing redshift
    (Figure~\ref{figure:vdisp_z}), the small-to-large
    ratio increases with decreasing $z$.  \label{figure:gsd_z}}
\end{figure}

With an understanding of the evolution of the physical conditions of
the dusty ISM in hand, we now turn to examining the impact on the
grain-grain collision velocity, and the subsequent dust grain size
distribution.  In Figure~\ref{figure:vdisp_z}, we present a boxen\footnote{Concretely, we compute $v_{\rm b}$ for every dust particle, averaging over each particle's grain size distribution and weighted by the grain mass per size bin.  At each snapshot redshift, the boxen plot shows the distribution of these particle velocities in all model galaxies, with each particle weighted by its total dust mass.  The nested boxes show successive quantiles of this mass-weighted distribution, with the central line denoting the median.} plot
of the distribution of mass-weighted dust velocities of every model
galaxy in our simulation sample.  As the typical ISM conditions
surrounding dust become more diffuse with decreasing redshift
(Figure~\ref{figure:dust_density_temp_z}), the interaction velocities
between dust particles rise (Equation~\ref{equation:vb_rhot}).  The
result is that significant numbers of dust particles become eligible
for shattering at $z\la 4$, which will impact the overall dust grain
size distribution.  (We note that the order of magnitude change in
velocities in Figure~\ref{figure:vdisp_z} may not match the order of
magnitude drop in $n_{\rm H}$ in
Figure~\ref{figure:dust_density_temp_z}.  This is due both to the
impact of temperature in Equation~\ref{equation:vb_rhot}, as well as
the self-regulating impact of grain sizes: as grain sizes become
smaller, they are better coupled with the turbulent flow, which lowers
grain-grain relative velocities.)

In Figure~\ref{figure:gsd_z}, we show the cosmic evolution of the
normalized dust grain size distributions of our model galaxies.  We
highlight in the grey shaded region the assumed size range of PAHs in
our model (noting that while we show the evolution of the size
distribution for all dust particles in our model, PAHs in our
framework exclusively are ultrasmall carbonaceous grains).  For
reference, we additionally show the \citet{mathis77a} constraint of
the grain size distribution in the Galaxy.  The top panel of
Figure~\ref{figure:gsd_z} shows the evolution of every model galaxy in
our sample over a broad range of redshifts ($z=0-10$), while the
bottom three panels bin the evolution of $3$ representative redshift
ranges (low redshift: $z<2$, intermediate redshift: $2\leq z \leq 6$,
and high redshift: $z>6$).  The transfer of power from large grains to
small grains in the distribution functions as cosmic time evolves is
immediately apparent.

\begin{figure*}
  \centering
  \includegraphics[scale=0.6]{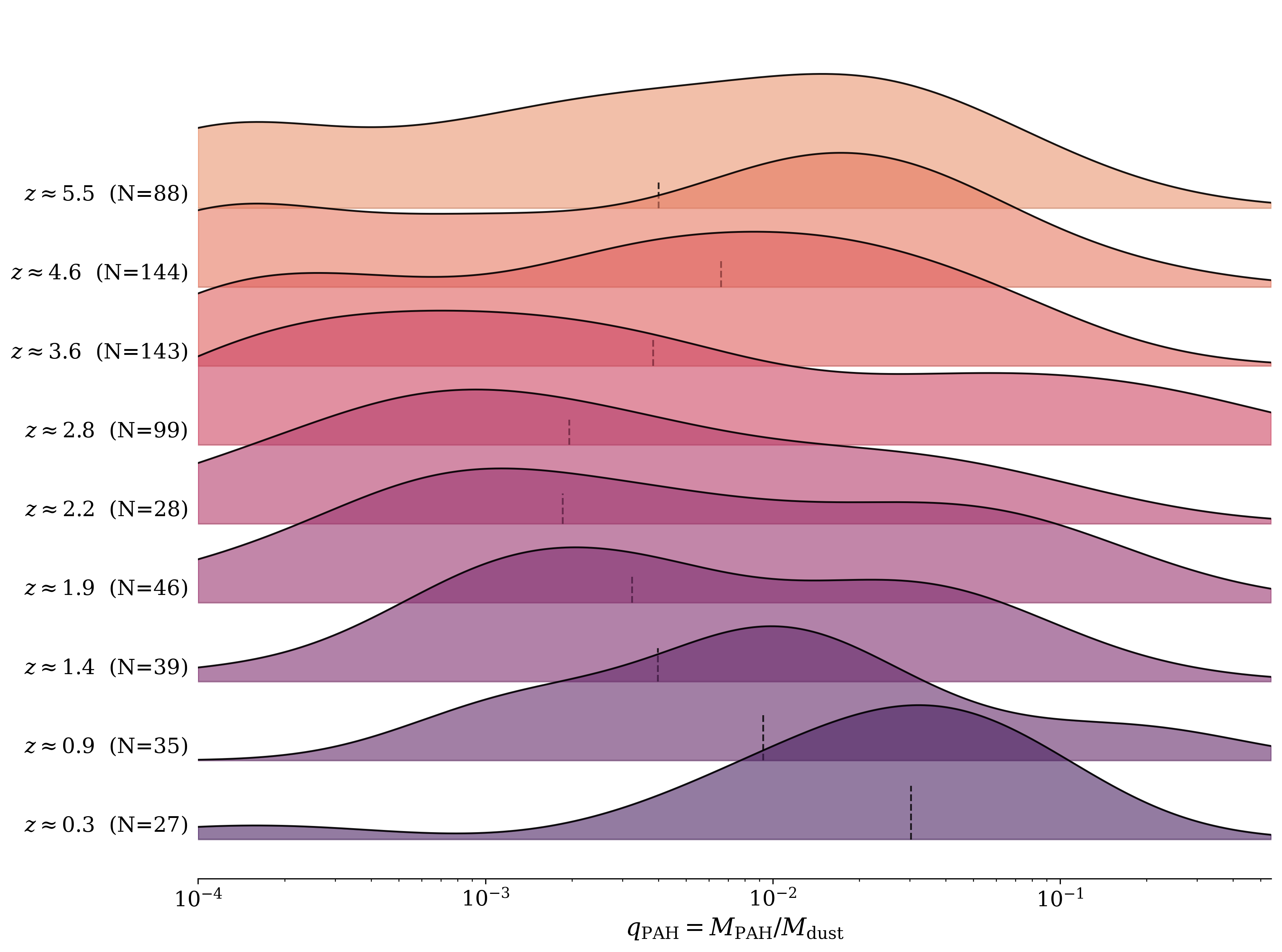}
  \caption{{\bf $q_{\rm PAH} \equiv M_{\rm PAH}/M_{\rm dust}$
      distributions for our model galaxies, binned by redshift.}  The
    median of each distribution is shown by a thin dashed vertical
    line, and the number of galaxies that comprise each distribution is noted by the redshift markers.  The median of the distribution follows the cosmic
      evolution of the gas density
      (cf. Figure~\ref{figure:dust_density_temp_z}).  As redshift
    decreases below $z \sim 4$, the mass fraction of diffuse ISM
    increases, resulting in larger dust interaction velocities.  This
    increases the grain-grain shattering events, and therefore
    fraction of ultrasmall dust grains.  The consequence of this is
    increased \qpah \ fractions between $z \approx 4 \rightarrow 0$,
    with median values of a few percent at $z \approx
    0$.  \label{figure:qpah_ridge}}
\end{figure*}

\begin{figure*}
  \centering
  \includegraphics[scale=0.65]{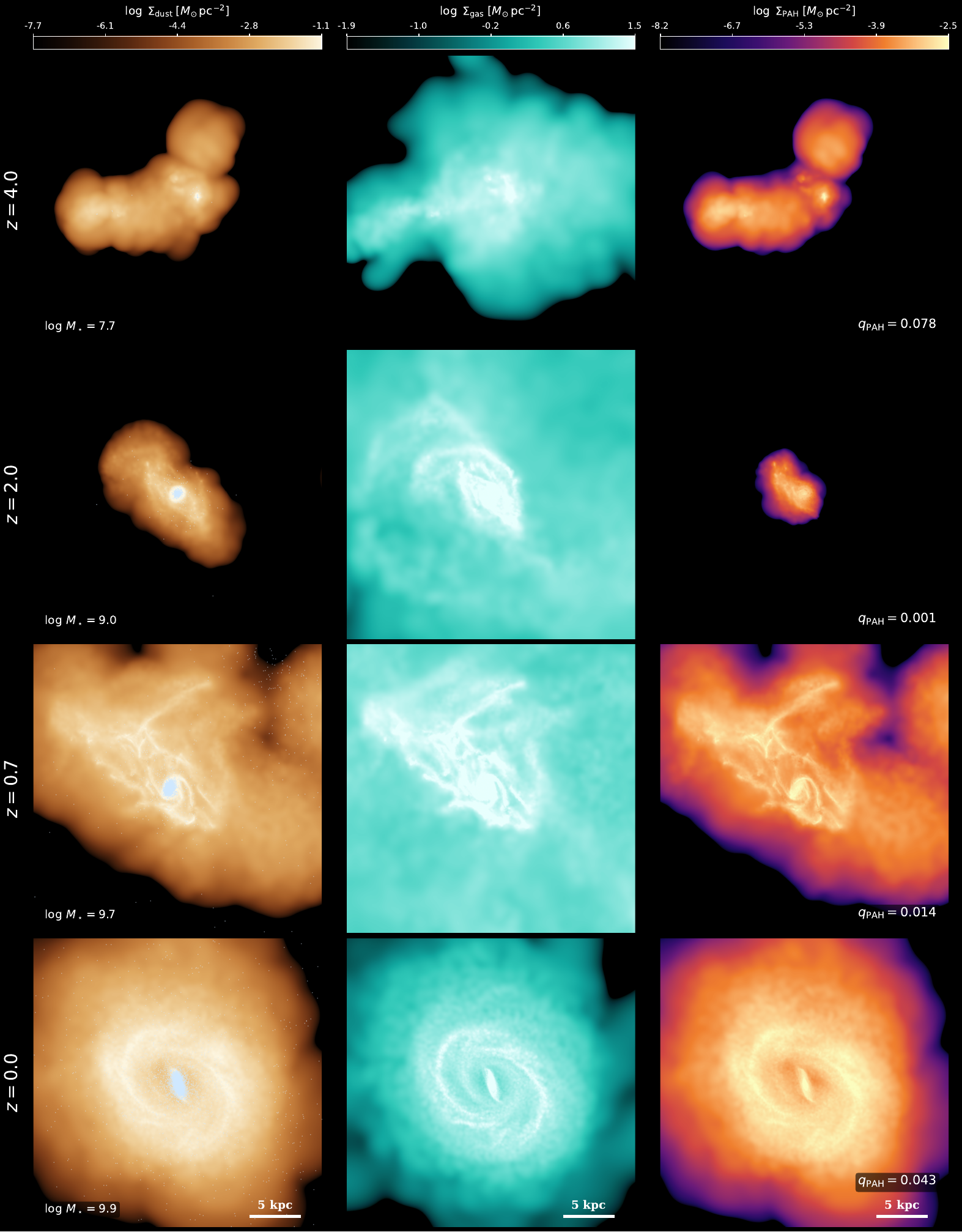}
  \caption{{\bf Morphological evolution of the dust, gas, and
        PAH surface density distributions for an example galaxy
        (galaxy 10407) from $z=4\rightarrow0$.}  Columns show, from
      left to right, the dust surface density $\Sigma_{\rm dust}$
      with the stellar distribution overplotted as individual star
      particles (points; as a collisionless component, the stars are
      shown discretely rather than as a continuous surface density
      field), gas
      surface density $\Sigma_{\rm gas}$, and PAH surface density
      $\Sigma_{\rm PAH}$, while rows correspond to $z = 4.0$, $2.0$,
      $0.75$, and $0.0$ (top to bottom).  Each column shares a common
      colorbar normalization.  At $z \sim 4$, the galaxy is compact
      and gas-rich, with PAHs confined to a small number of sites.  By
      $z \sim 0$, the galaxy has developed an extended disk, and PAHs
      trace the diffuse ISM spiral structure
      (cf.\ Figures~\ref{figure:dust_density_temp_z}--\ref{figure:qpah_fmol}).
      The global PAH mass fraction $q_{\rm PAH}$ (annotated in the
      right column) increases from $< 0.1\%$ at $z=2$ to $\sim 4\%$ at
      $z=0$.
      \label{figure:pah_morphology_evolution}}
\end{figure*}

Taken together, the evolution of gas densities
(Figure~\ref{figure:dust_density_temp_z}) results in an increase in
dust interaction velocities (Figure~\ref{figure:vdisp_z}), which
transforms the grain size distributions to include orders of magnitude
more small grains (Figure~\ref{figure:gsd_z}).  There are two key
consequences of these physical processes.  First, our models predict
an increase in the mass fraction of dust that is considered PAHs with
decreasing redshift\footnote{There is, of course, a key aspect of this
analysis that is missing: the {\it composition} of the dust grains:
PAHs are assumed in our model to be ultrasmall carbonaceous grains.
This said, this is answered by the fact that the fraction of
carbonaceous grains increases as redshift decreases (of course, not necessarily monotonically, depending on the nature of species-dependent growth in galaxies as a function of redshift.).  In our model,
the vast majority of carbonaceous grains are formed in AGB stars which
have a delay time from a star formation event from Type II SNe (which
produce primarily silicates in our model).  As a result, as the
Universe evolves and more stars move to the AGB, the fraction of
available carbonaceous grains simply increases.}.  We show this
explicitly in Figure~\ref{figure:qpah_ridge}, where we show the
relative distribution of $q_{\rm PAH} \equiv M_{\rm PAH}/M_{\rm dust}$
for all of our model galaxies, binned by redshift.  The thin dashed
vertical lines in each redshift bin denote the median in the
distribution.  As is clear, the fraction of PAHs follows the gas
physical conditions, interaction velocities, and grain size
distributions, with typical \qpah \ values at $z \sim 0$ broadly
matching the inferences from local observational constraints of a few
percent \citep[e.g.][]{aniano20a}.   We visualize this evolution directly in
Figure~\ref{figure:pah_morphology_evolution}, where we show the
projected dust, gas, and PAH surface density maps (with the
stellar distribution overplotted as individual star particles on the
dust maps) for galaxy 10407
at four epochs spanning $z = 4 \rightarrow 0$.  We choose galaxy
10407 as it is one of the $8$ galaxies in our sample evolved to
$z=0$, and its stellar mass, SFR, and gas-phase metallicity all lie
near the medians of that subset (Table~\ref{table:galaxy_sample}),
making it a representative low-redshift star-forming disc.  The spatial
distribution of PAHs evolves from compact and sparse at high redshift
to an extended structure tracing the diffuse spiral arms at $z = 0$.

\begin{figure}
  \includegraphics[scale=0.4]{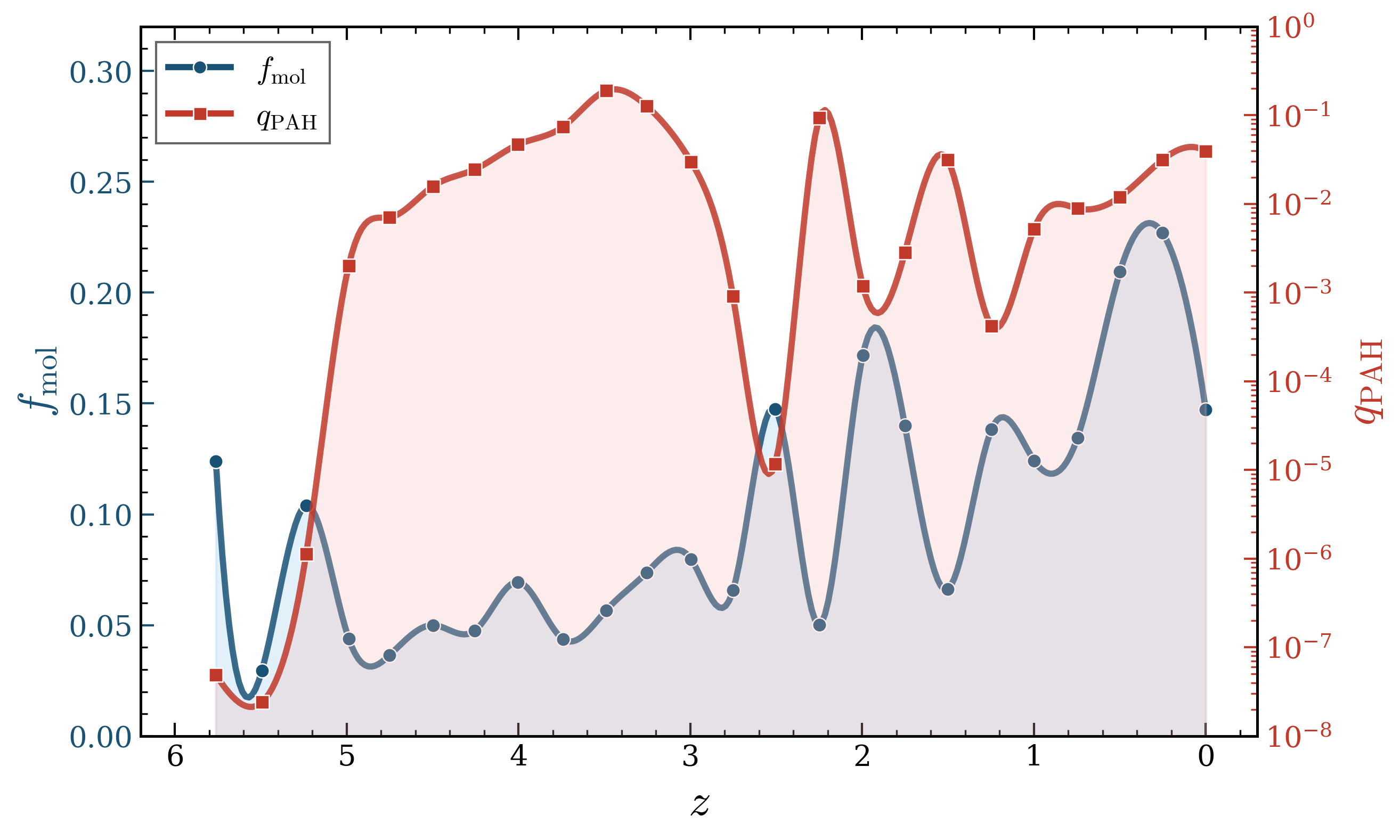}
  \caption{{\bf The molecular gas fraction and $q_{\rm PAH}$ evolve
      inversely over cosmic time.}  We show the redshift evolution of
    the molecular gas fraction $f_{\rm mol}$ (blue; left axis) and
    $q_{\rm PAH} \equiv M_{\rm PAH}/M_{\rm dust}$ (red; right axis)
    for an example galaxy (galaxy 10407) from our simulation sample.
    The two quantities track nearly inversely such that when the $f_{\rm mol}$ is low, $q_{\rm PAH}$ is elevated owing to increased grain-grain interaction velocities in the diffuse ISM (which increases shattering rates, and therefore, the fraction of PAH grains).    \label{figure:qpah_fmol}}
\end{figure}

Second, there is an implied inverse correlation between the dense gas
fraction and \qpah \ in our model galaxies.  We demonstrate this
explicitly in Figure~\ref{figure:qpah_fmol}, where we plot the time
evolution of the molecular gas fraction (as a proxy for the dense gas
fraction) and \qpah \ as a function of redshift for an example galaxy
(galaxy 10407).  The two operate almost entirely inversely: increased
diffuse gas fractions (lower $f_{\rm mol}$) result in increased PAH
mass fractions\footnote{There is of course a significant
  discrepancy in the dynamic range spanned by the $q_{\rm PAH}$
  variations, and $f_{\rm mol}$ variations.  This is because $f_{\rm
    mol}$ is ultimately a crude tracer of the true physical attribute
  at play, which is the average density of the gas.  Still, we choose
  to parameterize Figure~\ref{figure:qpah_fmol} in terms of the
  molecular fraction, as this will have important consequences for the
  PAH-CO relation in \S~\ref{section:pah_sfr}.} over the life of the
galaxy, owing to the fact that in our model, the main physical
  anti-correlation here is between the molecular fraction and
  interstellar velocity.

\subsection{The Destruction of PAHs}
Missing from our study of the emergence of PAHs in cosmological
simulations is an investigation of the main destruction pathways of
PAHs.  To study this, we perform a targeted numerical experiment with
a single test galaxy, galaxy 10407.  We run these numerical experiments down to $z=0$, and
perform three tests: (1) a model where destruction of dust grains by
SNe is turned off; (2) a model where thermal sputtering is turned off,
and (3) a model where coagulation is turned off.  The first two
numerical tests study the impact of dust destruction, while the latter
test investigates the potential impact of grain-grain coagulation as a
means of transforming ultrasmall dust grains into larger grains
(therefore conserving dust mass, but removing PAHs from the system).

\begin{figure}
  \includegraphics[scale=0.4]{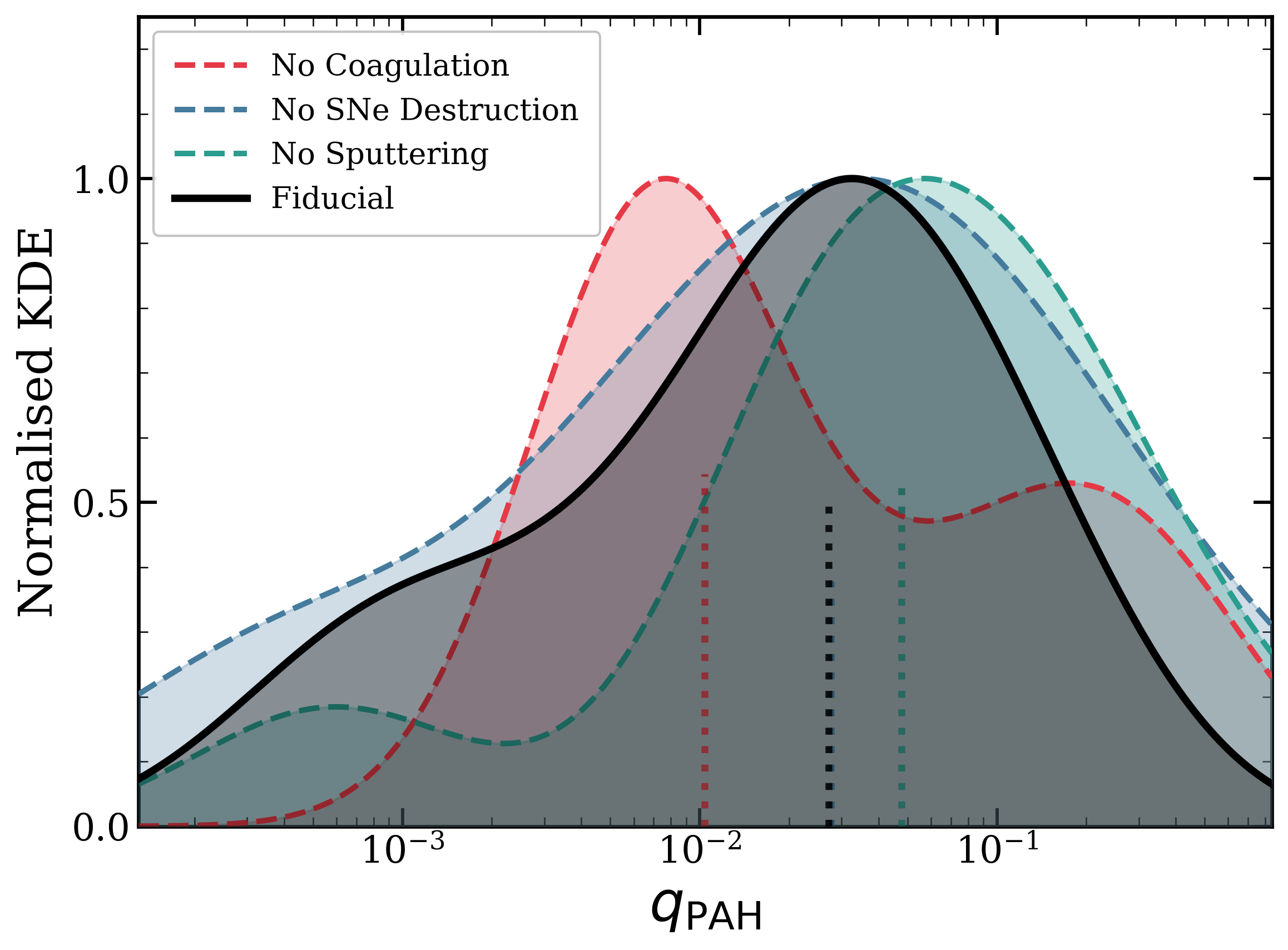}
  \caption{{\bf Quantitative comparison of impact of PAH destruction
      mechanisms in our simulations.}  Shown are histograms of $3$
    numerical experiments compared to a fiducial model galaxy from our
    simulation sample (galaxy 10407) between $z=0-6$.  The $3$ experiments include
    turning off the impact of coagulation, SNe destruction of grains,
    and thermal sputtering.  Thermal sputtering emerges as the
    dominant mechanism for destroying PAHs in our simulations, though
    the impact is modest (factor $\sim 2$ in $q_{\rm PAH}$).  It is important
    to note that we do not include the radiative destruction of PAHs
    or destruction near HII regions in our
    simulations.\label{figure:mpah_z_destruction}}
\end{figure}

In Figure~\ref{figure:mpah_z_destruction}, we plot histograms of all
of the \qpah \ values (for every saved snapshot) for the $3$ numerical
experiments, as well as our fiducial galaxy model with all standard
physics implemented.  The median of each distribution is given by a
vertical dashed line (with the fiducial model shown in black).
It is immediately clear that SNe destruction does not preferentially
impact PAHs in our models\footnote{In theory, PAHs exposed to
  supernovae are destroyed extremely effectively.  Following
  \citet{nozawa06a} and \citet{asano13b}, grains with $a<100 \AA$ are
  destroyed efficiently by supernovae shocks.  This said,
  SNe destruction is a high-efficiency, low-duty-cycle process: most
  PAHs are not exposed to a SNe, resulting in an overall low
  destruction rate.  In contrast, thermal sputtering operates
  continuously on all dust in hot gas, giving it a higher integrated
  impact, despite lower per-interaction efficiency.}, while turning
off coagulation actually {\it decreases} the PAH fraction.  This can
be understood by a decrease in the fraction of large grains (i.e., a
lack of conversion of small grains into large grains) available to
shatter and form PAHs. In our simulations, coagulation is a
  relatively inefficient mechanism for removing PAHs via simple
  transformation from ultrasmall$\rightarrow$ large grains.
Ultimately, in our numerical tests, turning off thermal sputtering has
the largest impact on \qpah, suggesting that it is the dominant
mechanism for destroying PAHs in our simulations.  Still, the impact
is modest, shifting the median in the distribution function from the
fiducial run by a factor $\sim 2$.  It is important to note that in
our models, we do not include either the radiative dissociation of
small grains, or the potential destruction within HII regions, both of
which may be important physical mechanisms in governing the PAH
destruction rate in the ISM of galaxies \citep{egorov23a}.  We
therefore hesitate to make strong statements regarding the dominant
source of PAH destruction in galaxies.

\section{Part II: The Luminous Properties of PAHs over Cosmic Time}
\label{section:luminous_properties}
In Part I of this paper, we developed a model for the physical
formation of PAHs in cosmological simulations of galaxy evolution.  In
Part II, we expand this analysis to the {\it luminous} properties of
PAHs in these galaxies.  As a reminder (\S~\ref{section:powderday}), we have
processed each of our model snapshots through the {\sc powderday} dust
radiative transfer package where the photon heating rate of PAHs is
computed.  Then, by tying these models to the single photon
approximation calculations of \citet{richie25a}, we arrive at the
emergent PAH spectrum for each model galaxy (see
Figure~\ref{figure:sed_decomposition} for representative {\sc pahfit}
decompositions of the simulated mid-IR spectra).  We now expand this
analysis to investigate the consequences of our model on the
rest-frame mid-IR spectrum of galaxies over cosmic time.  In particular, we focus on the PAH-metallicity relationship, the PAH-star formation rate (SFR) relationship, and the PAH-molecular gas relationship in galaxies.

\begin{figure*}
  \centering
  \includegraphics[scale=0.4]{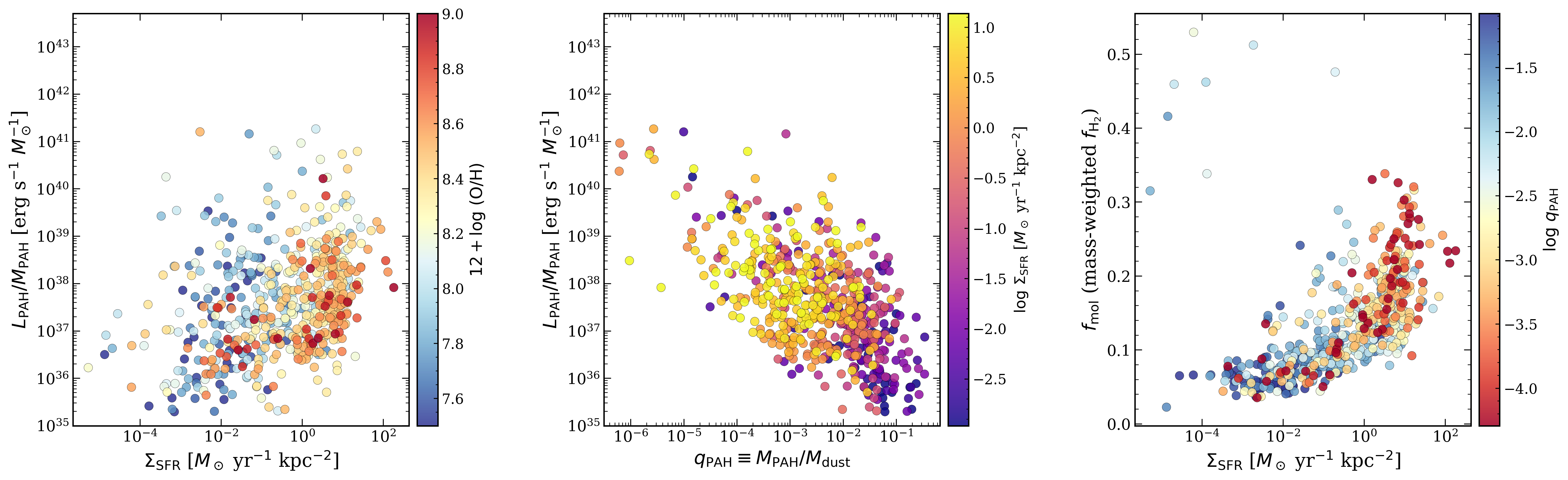}
  \caption{{\bf The PAH Light to Mass Ratio correlates with
      $G_0$, and anti-correlates with $q_{\rm PAH}$ in our models}.
    {\it Left:} Correlation between PAH $L/M$ ratio with $\Sigma_{\rm
      SFR}$, serving as a proxy for $G_0$; each point represents a
    simulation galaxy snapshot, and is color coded by its metallicity.
    {\it Middle:} The PAH $L/M$ ratio vs the PAH mass
    fraction $q_{\rm PAH}$, color coded by its $\Sigma_{\rm SFR}$.
    Higher $\Sigma_{\rm SFR}$ galaxies more easily excite PAHs,
    resulting in a higher light-to-mass ratio.  At the same time,
    there is an anti-correlation with $q_{\rm PAH}$.  This owes to the fact
    that higher $\Sigma_{\rm SFR}$ galaxies have higher molecular gas
    fractions, and therefore lower typical collision velocities
    between dust grains, resulting in lower PAH production (cf.  Part I of this paper).  {\it Right:} Correlation between $\Sigma_{\rm SFR}$ and $f_{\rm mol}$, color coded by $q_{\rm PAH}$.
    \label{figure:lpah_mpah_sigma_sfr}}
\end{figure*}

\subsection{The Cosmic Evolution of $L_{\rm PAH}$ in Galaxies}
\label{section:lpah_mpah}
As described in \S~\ref{section:helena}, PAH emission (for
  ultrasmall grains that are spatially removed from intense sources of
  radiation sources such as H~{\sc ii} regions) operates in the single-photon
regime: each absorbed UV photon vibrationally excites a PAH molecule,
which cools by radiating in the classical MIR bands.  The key
consequence of this physics is that the total emitted power scales
linearly with the photon absorption rate.  We now derive this scaling
and discuss its implications for interpreting PAH observations.

As a reminder, the core quantity in the \citet{richie25a} model is the
basis spectrum: the time-averaged power emitted per unit wavelength by a grain of size $a$ 
during one complete cooling event from peak temperature $T_{\rm peak}$
back to the ground state is given by (a slight recast of Equation~\ref{equation:spa_cooling}),
\begin{equation}
  \tilde{p}_{\lambda_{\rm em}}(\lambda_{\rm abs})
  \;=\;
  \frac{4\pi}{t_{\rm max}}
  \int_0^{t_{\rm max}}
  B_{\lambda_{\rm em}}\!\bigl(T(t)\bigr)\,C_{\rm abs}(\lambda_{\rm em})\,dt,
  \label{eq:basis_spectrum}
\end{equation}
for $\lambda_{\rm em} \geq \lambda_{\rm abs}$, and $\tilde{p}_{\lambda_{\rm em}}(\lambda_{\rm abs}) = 0$ otherwise,
where $\tilde{p}_{\lambda_{\rm em}}(\lambda_{\rm abs})$ is the basis spectrum (power emitted per unit wavelength for a PAH absorbing a single photon at wavelength $\lambda_{\rm abs}$), $\lambda_{\rm em}$ is the emitted wavelength, $t_{\rm max}$ is the duration of a single cooling event, $B_{\lambda_{\rm em}}$ is the Planck function, $C_{\rm abs}$ is the grain
absorption cross-section, and $T(t)$ is the grain temperature as a function of time, obtained by solving the radiative cooling
equation.
The basis spectrum encodes which mid-IR bands receive power for a given
absorbed photon energy.
Weighting the basis spectrum by the rate of photon absorption from each wavelength
channel $\lambda_{\rm abs}$ gives the full emission spectrum \citep{richie25a}:
\begin{equation}
  p_{\lambda_{\rm em}}(\lambda_{\rm abs})
  \;=\;
  \frac{\displaystyle
        \int_{\lambda_{\rm abs}}^{\lambda_{\rm abs}(1+\Delta)}
        c\,u_\lambda(\lambda)\,C_{\rm abs}(\lambda)\,d\lambda}
       {\displaystyle
        \int_{\lambda_{\rm em}}
        \tilde{p}_{\lambda_{\rm em}}(\lambda_{\rm abs})\,d\lambda_{\rm em}}
  \;\tilde{p}_{\lambda_{\rm em}}(\lambda_{\rm abs}),
  \label{eq:normalized_spectrum}
\end{equation}
where $p_{\lambda_{\rm em}}(\lambda_{\rm abs})$ is the full emission spectrum per grain, $c$ is the speed of light, $\Delta$ is the (dimensionless) fractional width of the logarithmic absorption wavelength bin, and $u_\lambda$ is the radiation field energy density.  The
numerator is the absorbed power in a narrow wavelength bin around
$\lambda_{\rm abs}$, while the denominator represents energy
normalization.  Crucially, the numerator scales linearly with
$u_\lambda \propto \Gzero$, while the denominator depends only on
grain properties ($\Gzero$ is the strength of the FUV interstellar radiation field, normalized to the \citet{habing68a} value).  Integrating $p_{\lambda_{\rm em}}$ over all
emission wavelengths and summing over all absorbed-photon channels
therefore gives the total single-grain PAH power as simply the rate of
UV energy absorption (see Appendix~\ref{appendix:lpah_derivation} for
the step-by-step derivation):
\begin{equation}
\begin{split}
  \ell_{\rm PAH}
  \;\equiv\;& \int p_{\lambda_{\rm em}}\,d\lambda_{\rm em}
  \;=\;
  \int C_{\rm abs}(\lambda)\,c\,u_\lambda\,d\lambda \\
  &\propto\;
  \pi a^{2}\,\Qabs^{\rm UV}\,\Gzero.
\end{split}
  \label{eq:lpah_grain}
\end{equation}
Here, $\ell_{\rm PAH}$ is the total IR power emitted by a single PAH grain, $a$ is the grain radius, and $\Qabs^{\rm UV}$ is the UV absorption efficiency (i.e., the dimensionless absorption cross-section $C_{\rm abs}/\pi a^2$ averaged across the FUV).
Summing over a population of PAH grains of total mass $\Mpah$ and using
$\Kappa_{\rm UV}^{\rm PAH} \equiv 3\Qabs^{\rm UV}/(4\,a\,\rho_{\rm gr})$ as the
UV absorption opacity per unit PAH mass (see Appendix~\ref{appendix:lpah_derivation}),
with $\rho_{\rm gr}$ the internal grain density,
this gives
\begin{equation}
 \frac{\Lpah}{\Mpah} \;\propto\; \Kappa_{\rm UV}^{\rm PAH}\,\Gzero.
  \label{equation:lpah_mpah_scaling}
\end{equation}
Equation~\ref{equation:lpah_mpah_scaling} simply reflects the fact that PAH emission traces the  absorbed photon rate
rather than a grain temperature
\citep{desert90a,draine11a}, and is confirmed observationally by
\citet{li20b}.  In
Figure~\ref{figure:lpah_mpah_sigma_sfr}, we show this
relationship from our simulations by comparing the $L_{\rm
  PAH}/M_{\rm PAH}$ ratio vs $\Sigma_{\rm SFR}$ ($L_{\rm PAH}$ is computed by integrating over the {\sc pahfit} fits to the mid-IR spectrum, while $\Sigma_{\rm SFR}$ is computed as the total
SFR divided by $r_{\rm half}^2$): the correlation is clear. 

At the same time, there is an inverse correlation in our models
between the PAH light to mass ratio, and the PAH mass fraction,
$q_{\rm PAH}$.  This is demonstrated explicitly in the middle panel of
Figure~\ref{figure:lpah_mpah_sigma_sfr}, which shows this inverse
relationship, color-coded by $\Sigma_{\rm SFR}$.  The origin of this
inverse relationship is rooted in the fact that high $\Sigma_{\rm
  SFR}$ galaxies (which have the highest intensity of
  PAH-exciting photons) tend to have higher molecular gas fractions,
and therefore lower fractions of the diffuse low-density gas that is
conducive for grain-grain shattering in our model
(Equation~\ref{equation:vb_rhot}).  In the right hand panel of
Figure~\ref{figure:lpah_mpah_sigma_sfr}, we confirm this by showing
the relationship between the molecular gas fraction in our model
galaxies, and the galaxy $\Sigma_{\rm SFR}$.  Taken together,
Figure~\ref{figure:lpah_mpah_sigma_sfr} paints a picture in which the
PAH light to mass ratio is decreasing with $q_{\rm PAH}$: in other
words, based on the local physical conditions, the mass fraction of
dust in PAHs may not linearly translate to the light it produces.

\begin{figure*}
  \centering
  \includegraphics[scale=0.4]{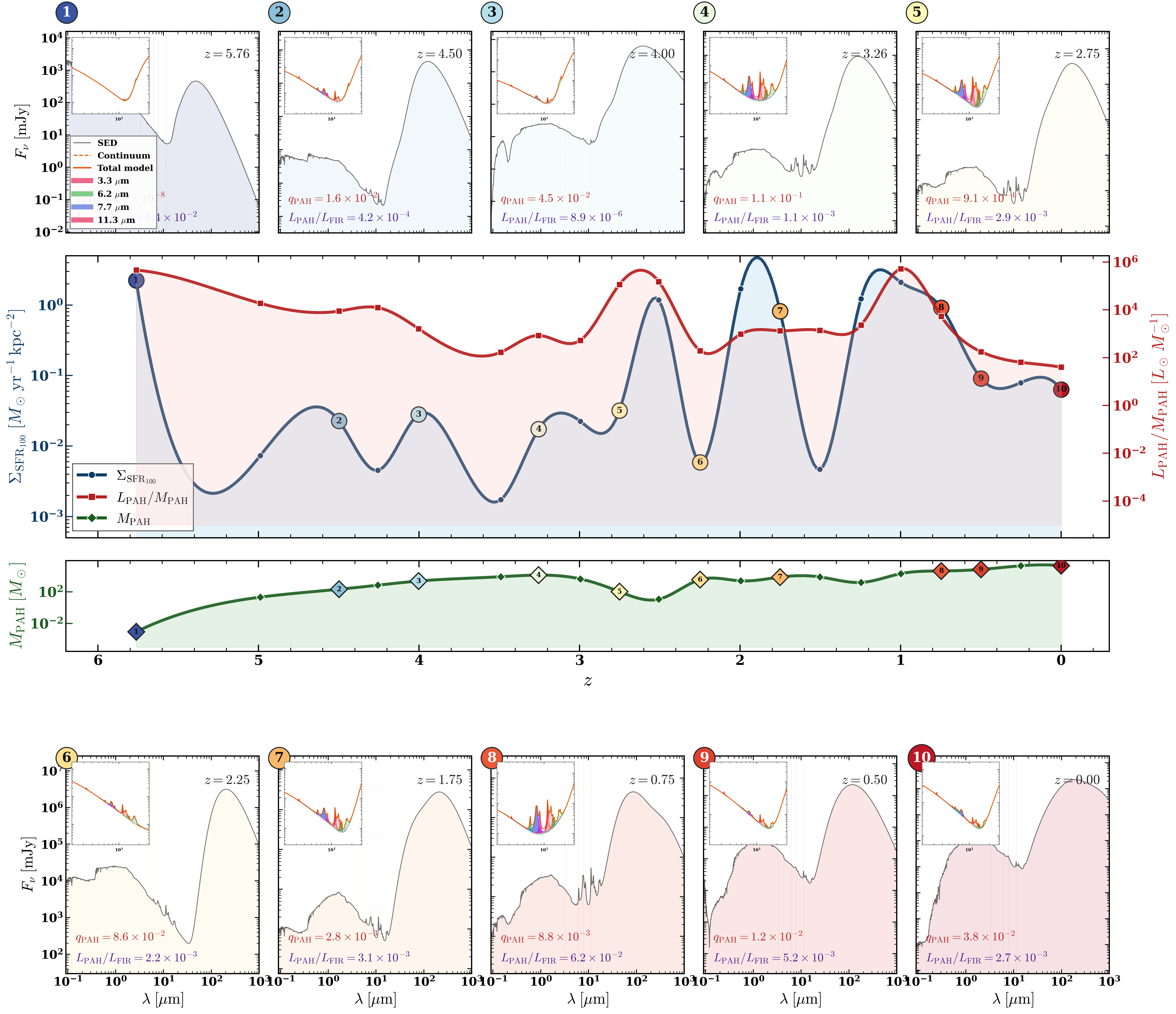}
  \caption{{\bf The co-evolution of $\Sigma_{\rm SFR}$, the PAH light
      to mass ratio, PAH mass, and the broadband SED for a single
      representative galaxy across cosmic time.} {\it Upper center:}
    The evolution of $\Sigma_{\rm SFR}$ (blue; left axis) and $L_{\rm
      PAH}/M_{\rm PAH}$ (red; right axis) from $z \sim 6$ to $z = 0$,
    with numbered circles marking $10$ individual snapshots.  {\it
      Lower center:} The evolution of $M_{\rm PAH}$ (green), which
    grows largely monotonically over cosmic time.  {\it Top and bottom
      panels:} Broadband SEDs corresponding to each numbered snapshot,
    with inset panels zooming into the mid-IR and showing the {\sc
      pahfit} decomposition of the PAH features (colored
    curves). Within each inset panel is both the observable
      $L_{\rm PAH}/L_{\rm FIR}$, as well as the physical \qpah \ for
      that snapshot.  While $M_{\rm PAH}$ increases steadily, $L_{\rm
      PAH}/M_{\rm PAH}$ fluctuates dramatically in concert with
    $\Sigma_{\rm SFR}$: during bursts of star formation, the radiation
    field drives $L_{\rm PAH}/M_{\rm PAH}$ upward, producing bright
    PAH emission features relative to the continuum.  During more
    quiescent phases, the PAH features are relatively subdued despite
    a steadily growing PAH mass.}
  \label{figure:sed_gallery}
\end{figure*}

To illustrate this interplay concretely, in
  Figure~\ref{figure:sed_gallery} we trace the evolution of a single
  representative galaxy across cosmic time. The upper center panel shows the
  co-evolution of $\Sigma_{\rm SFR}$ and $L_{\rm PAH}/M_{\rm PAH}$,
  while the lower center panel shows the evolution of $M_{\rm PAH}$
  itself; individual numbers mark snapshots whose broadband SEDs are
  displayed in the surrounding panels (with cutouts zooming in on the
  mid-IR portion of the SED, highlighting the {\sc pahfit}
  fits).  While $M_{\rm PAH}$ grows largely monotonically
  with time, $L_{\rm PAH}/M_{\rm PAH}$ fluctuates dramatically in
  response to changes in $\Sigma_{\rm SFR}$: during periods of
  elevated star formation activity, the radiation field intensifies
  and the PAH features brighten relative to the underlying dust
  continuum, driving $L_{\rm PAH}/M_{\rm PAH}$ upward. Conversely,
  during more quiescent phases, the weakened radiation field produces
  markedly subdued PAH emission despite a steadily growing underlying
  PAH mass.

\begin{figure}
  \centering
  \includegraphics[scale=0.4]{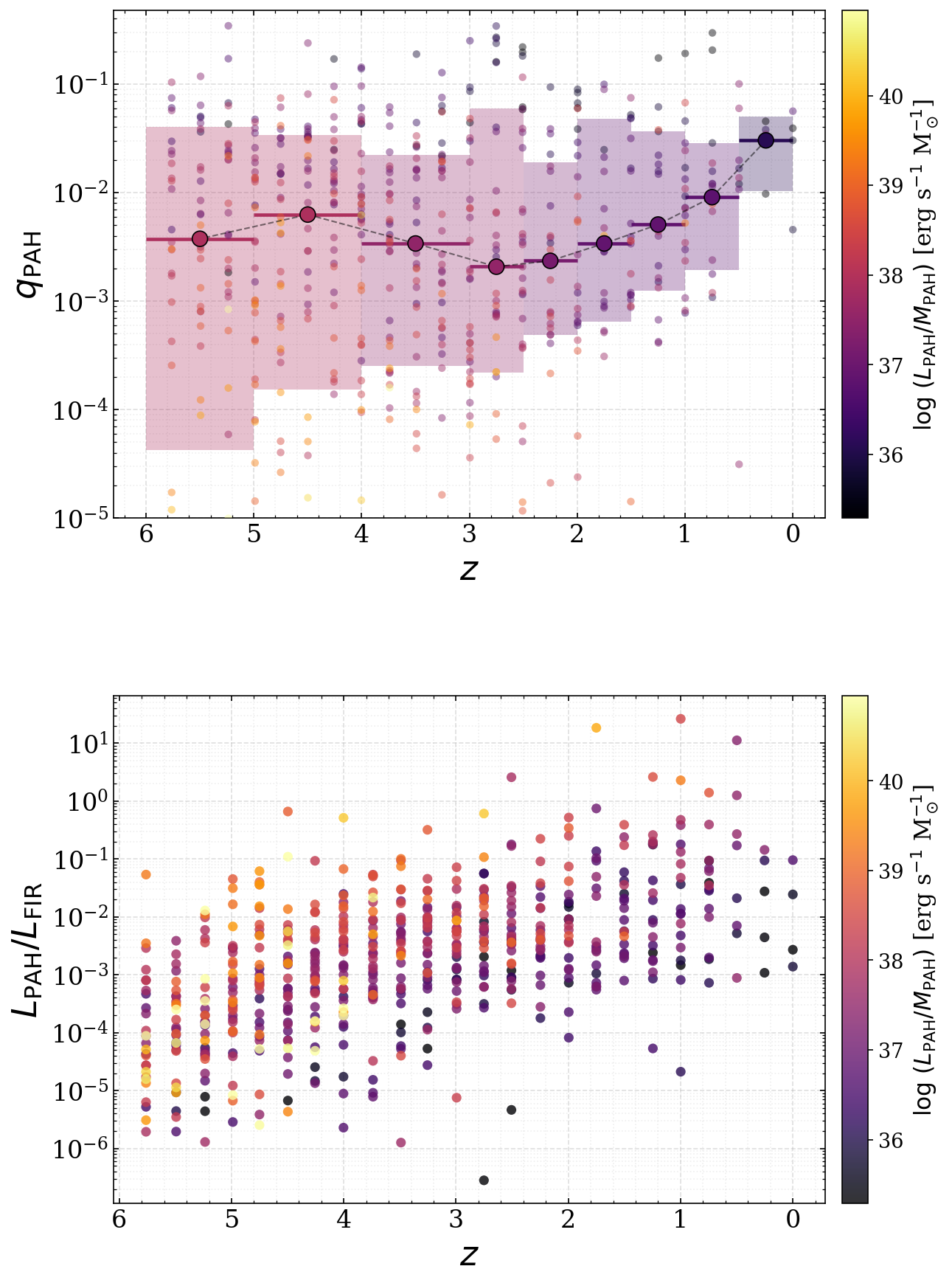}
  \caption{{\bf The redshift evolution of $q_{\rm PAH}$ and $L_{\rm
        PAH}/L_{\rm FIR}$ for our model galaxies do not perfectly
      correlate, owing to trends in the PAH light to mass ratio with
      $q_{\rm PAH}$.} {\it Top:} Redshift evolution of $q_{\rm PAH}$ for every model
    galaxy snapshot, color-coded by the galaxy PAH light to mass
    ratio.  {\it Bottom:} Same, but for $L_{\rm PAH}/L_{\rm FIR}$.
    Owing to the interdependency between $q_{\rm PAH}$, $\Sigma_{\rm SFR}$,
    and the molecular gas fraction in galaxies
    (cf.~Figure~\ref{figure:lpah_mpah_sigma_sfr}), $q_{\rm PAH}$ and
    $L_{\rm PAH}/L_{\rm FIR}$ do not perfectly track each other.}
  \label{figure:qpah_z}
\end{figure}

Ultimately this inverse relationship between the light to mass ratio
and $q_{\rm PAH}$ demonstrated in
Figure~\ref{figure:lpah_mpah_sigma_sfr} results in different
evolutionary behavior when comparing the physical \qpah \ vs $z$ and
the observed quantity $L_{\rm PAH}/L_{\rm FIR}$ vs $z$.  In
Figure~\ref{figure:qpah_z}, we show the evolution of these two
quantities with redshift for our model galaxies.  The top panel of
Figure~\ref{figure:qpah_z} is simply a re-casting of the distribution
functions plotted in Figure~\ref{figure:qpah_ridge}, while the bottom
panel of Figure~\ref{figure:qpah_z} transforms this to observed
space\footnote{Figures~\ref{figure:lpah_mpah_sigma_sfr}
  and ~\ref{figure:qpah_z} beg the question: How well does \qpah \ map
  to the observable quantity $L_{\rm PAH}/L_{\rm FIR}$?  A quick
  inspection of the numerical values inset in each SED panel in
  Figure~\ref{figure:sed_gallery} suggests that $L_{\rm
    PAH}/L_{\rm FIR}$ may only trace \qpah \ loosely.  This can owe
  both to the hardness of the radiation field (i.e.  lower energy
  photons can heat large dust grains), as well as the differential
  star-dust geometry between PAH grains and large grains.  We defer a
  full investigation of the mapping of PAH mass fractions to
  observable signatures, as well as the ability for modern fitting
  software to accurately derive \qpah \ to a forthcoming study.}.


Having established that the PAH light-to-mass ratio scales with
$\Gzero$ while anti-correlating with $q_{\rm PAH}$, we now turn to the
consequences of this framework for three key observational trends: the
PAH-metallicity relationship, the $L_{\rm PAH}$-SFR relationship, and
the $L_{\rm PAH}-M_{\rm mol}$ relationship in galaxies.  A central
implication of the results in this section is that the observed
luminous signature of PAHs encodes not only the PAH mass fraction, but
also the local radiation field intensity, which is a distinction that
will prove critical in interpreting the aforementioned scaling
relations.

\subsection{The PAH-Metallicity Relationship in Central Galaxies}
\label{section:pzr}
As discussed in \S~\ref{section:introduction}, the observed suppression of PAH emission at low
metallicities has been well documented from the {\it Spitzer} era
through JWST
\citep{engelbracht08a,aniano20a,shim23a,whitcomb24a,shivaei24a},
though the physical origin of the PZR remains debated.

In our model interpretation, the origin of the PZR can be explained by
the core physical concepts discussed in Part I: as galaxies grow over
cosmic time, a larger mass fraction of their ISM shifts from dense
molecular clouds to a diffuse state.  As dust moves from dense gas to
diffuse gas, its shattering rates go up, and the PAH mass fraction
increases.  At the same time, the metallicities of galaxies naturally
grow with cosmic time.  Critically, as discussed in
\S~\ref{section:lpah_mpah}, the {\it observed} PZR reflects not only
this physical growth in \qpah, but also the evolving radiation field
conditions that modulate the PAH light-to-mass ratio.

\begin{figure}
  \includegraphics[scale=0.5]{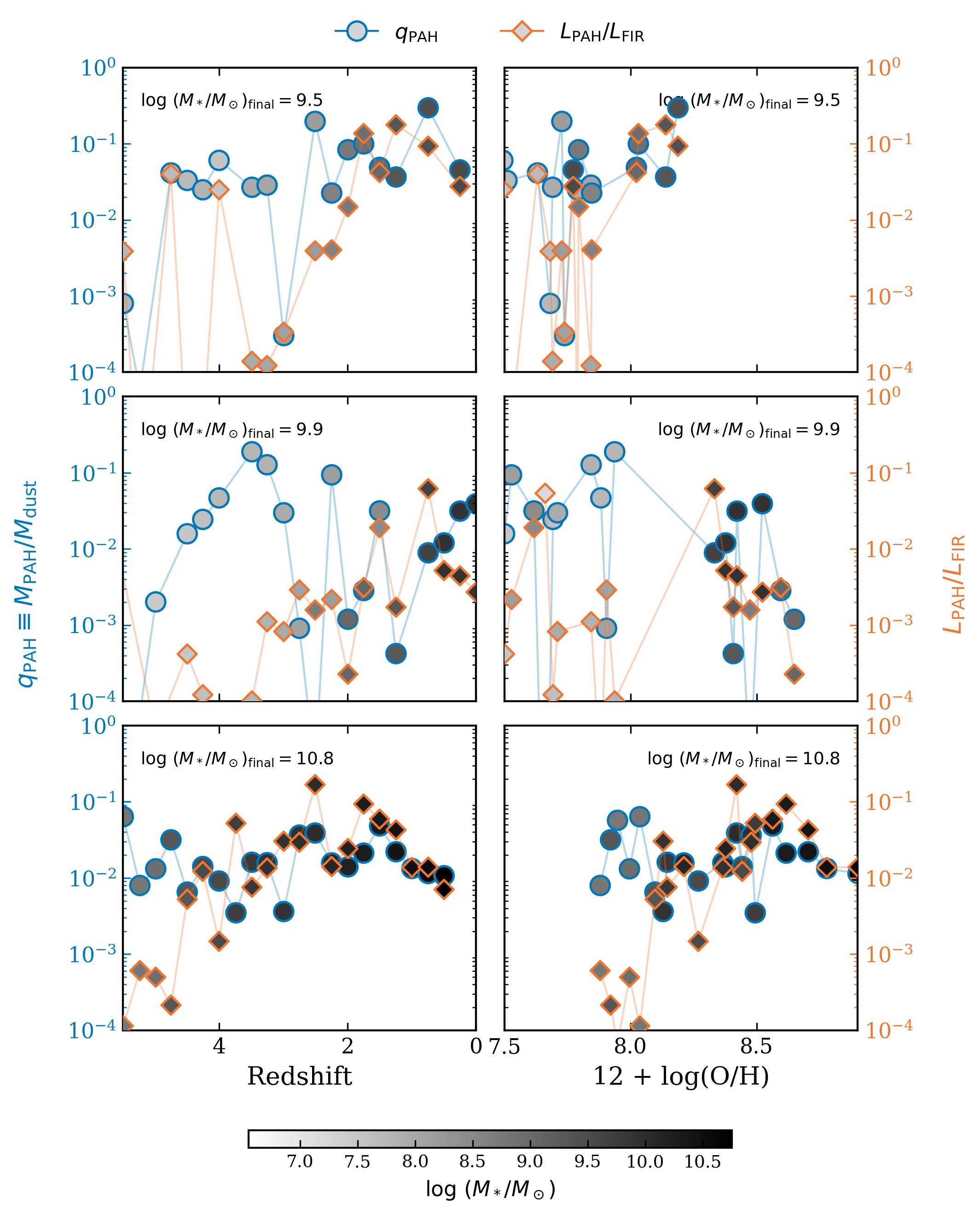}
  \caption{{\bf Evolution of $q_{\rm PAH}$ (circles) and $L_{\rm
        PAH}/L_{\rm FIR}$ (diamonds) with redshift (left) and
      metallicity (right) for $3$ individual galaxies spanning a range
      of final stellar masses ($\log(M_*/M_\odot)_{\rm final} = 9.5$,
      $9.9$, and $10.8$; top to bottom).}  Points are color-coded
      in grey scale by galaxy stellar mass, and the edge colors denote
      whether the symbol refers to the redshift evolution of \qpah
      \ (left) or $L_{\rm PAH}/L_{\rm FIR}$ (right).  While both
    quantities generally increase toward lower redshifts and higher
    metallicities, discrepancies arise from variable radiation fields
    (cf.~\S~\ref{section:lpah_mpah}).  Significant temporal
    fluctuations in both quantities reflect feedback-driven
    ``breathing modes'' in the ISM that modulate the diffuse gas
    fraction and, consequently, the grain shattering rate.
    \label{figure:pzr_lpah_lfir_z_Z}}
\end{figure}

We first quantify this effect in
Figure~\ref{figure:pzr_lpah_lfir_z_Z}, where we show the evolution of
\qpah \ vs both redshift and metallicity for three model galaxies
over a range of final galaxy stellar masses.  The left column of
Figure~\ref{figure:pzr_lpah_lfir_z_Z} shows the evolution of the
physical \qpah \ (blue) vs $z$ while the right column shows the same
vs $Z$.  We additionally show the evolution of $L_{\rm PAH}/L_{\rm
  FIR}$ (orange) vs $z$ and $Z$ as the most straightforward
observational manifestation of $q_{\rm PAH}$.  A number of salient
points arise in Figure~\ref{figure:pzr_lpah_lfir_z_Z}.  First, there
is a general rise in both the physical \qpah \ with decreasing
redshift, and the observed quantity, $L_{\rm PAH}/L_{\rm FIR}$, though
the two do not operate in lock step owing to variable interstellar radiation fields.  Second, while
the evolution in \qpah \ with $z$ (and therefore, $Z$) rises in
general, there are significant temporal fluctuations.  Even if the
general trend is to increase the fraction of diffuse gas with
decreasing redshift (Figure~\ref{figure:dust_density_temp_z}), there
are fluctuations in the diffuse gas mass fraction owing to the nature
of the feedback model.  Explicit feedback models such as {\sc fire}
and {\sc smuggle} experience ``breathing-modes'' in the ISM, where ISM
densities fluctuate rapidly as feedback blows gas out, it cools, forms
stars, and is blown out again \citep[e.g.][]{muratov15a,narayanan15a,mcclymont26a}.  These fluctuations result in
stochasticity in the small to large ratio for an individual galaxy,
and as a result, the \qpah \ evolution.  Third, there is an inherent
dispersion in the mass-metallicity relationship
(Figure~\ref{figure:behroozi}), with up to an order of magnitude dynamic
range in stellar mass at a fixed metallicity ultimately resulting in
an inherent dispersion in the $q_{\rm PAH}$-$Z$ relationship.

\begin{figure*}
  \centering
  \includegraphics[scale=0.4]{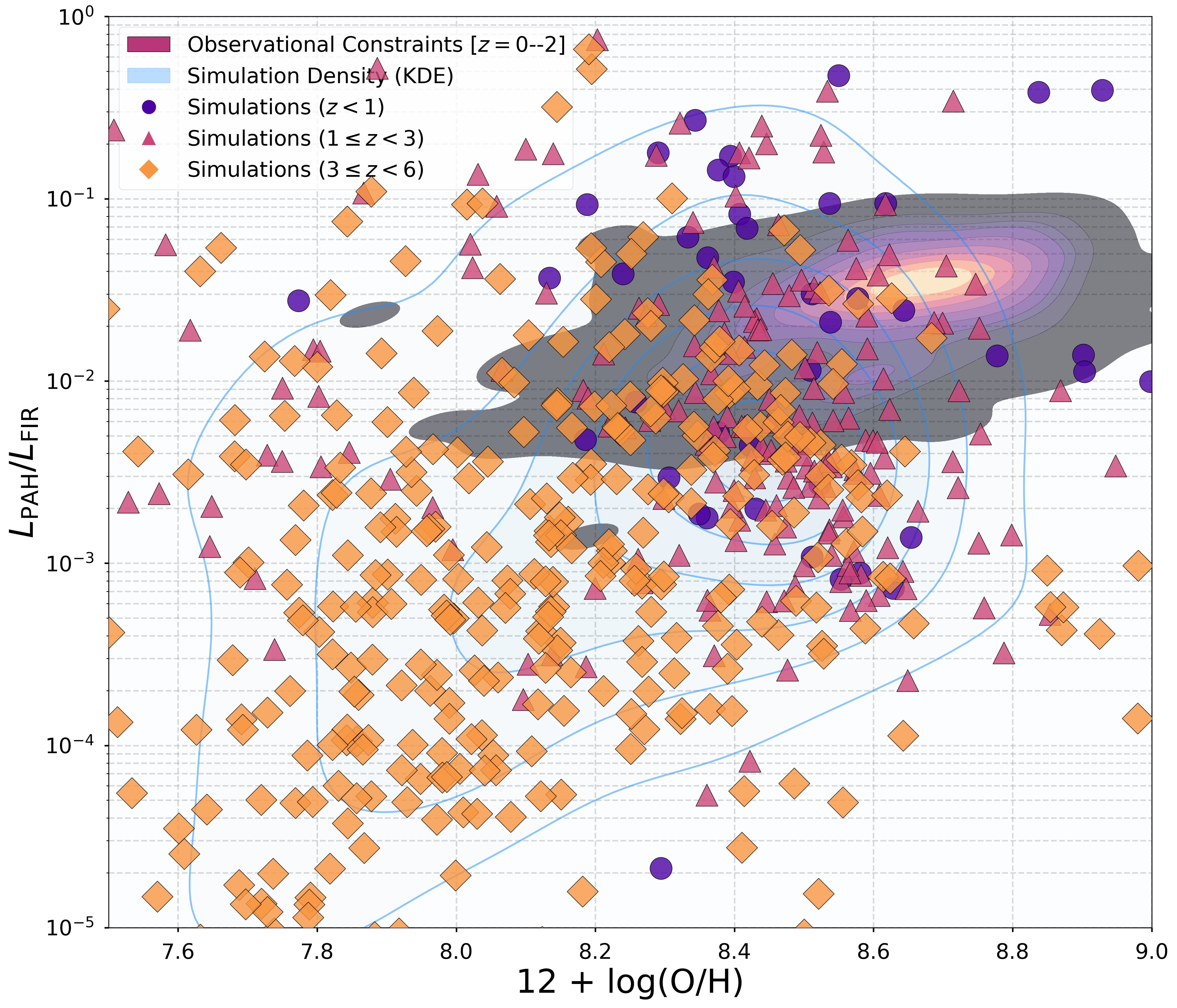}
  \caption{{\bf Modeled PAH-metallicity
      relationship (PZR) [squares] compared to observations (contours).} Model galaxy snapshots are binned and  color-coded by redshift, and the observational constraints are a compilation from \citet{aniano20a}, \citet{shim23a} and \citet{shivaei24a} aggregated
    across $z=0$-$2$ (contours).  We additionally show kernel density estimate contours of the simulation points in light blue. The model PZR arises naturally from
    the increasing fraction of diffuse ISM with cosmic time, which
    enhances grain-grain shattering and elevates $q_{\rm PAH}$.  The
    conversion from $q_{\rm PAH}$ to $L_{\rm PAH}/L_{\rm FIR}$ is
    non-trivial: elevated physical $q_{\rm PAH}$ values at high-$z$
    can translate to lower $L_{\rm PAH}/L_{\rm FIR}$ ratios owing to
    the dependence of the PAH light-to-mass ratio on the radiation
    field (cf.~\S~\ref{section:lpah_mpah}).
    \label{figure:pzr}}
\end{figure*}

The upshot of the analysis surrounding
Figure~\ref{figure:pzr_lpah_lfir_z_Z} is that our model predicts a
global, integrated relationship between the $L_{\rm PAH}/L_{\rm FIR}$
ratio and galaxy metallicity, though with a significant dispersion.
We demonstrate this explicitly in Figure~\ref{figure:pzr}, where we
show the model PZR (squares), color-coded by redshift, compared
against a compilation of literature observational constraints
(contours).  We have aggregated observational constraints across a
diverse range of redshifts ($z=0-2$) for this comparison, and draw
from the published datasets of \citet{aniano20a}, \citet{shim23a} and
\citet{shivaei24a}.  While we do not fit \qpah \ from SED fitting
methods, we cast Figure~\ref{figure:pzr} in observational space by
computing \qpah \ as $L_{\rm PAH}/L_{\rm FIR}$, which is a reasonable approximation for observational \qpah \ fitting methods. 
Figure~\ref{figure:pzr} manifests the intuition that we've built thus
far from these simulations that the relative power in PAH bands
compared to that of larger dust grains increases with increasing
metallicity, as the shattering rates increase.


\begin{figure*}
\centering
\includegraphics[scale=0.35]{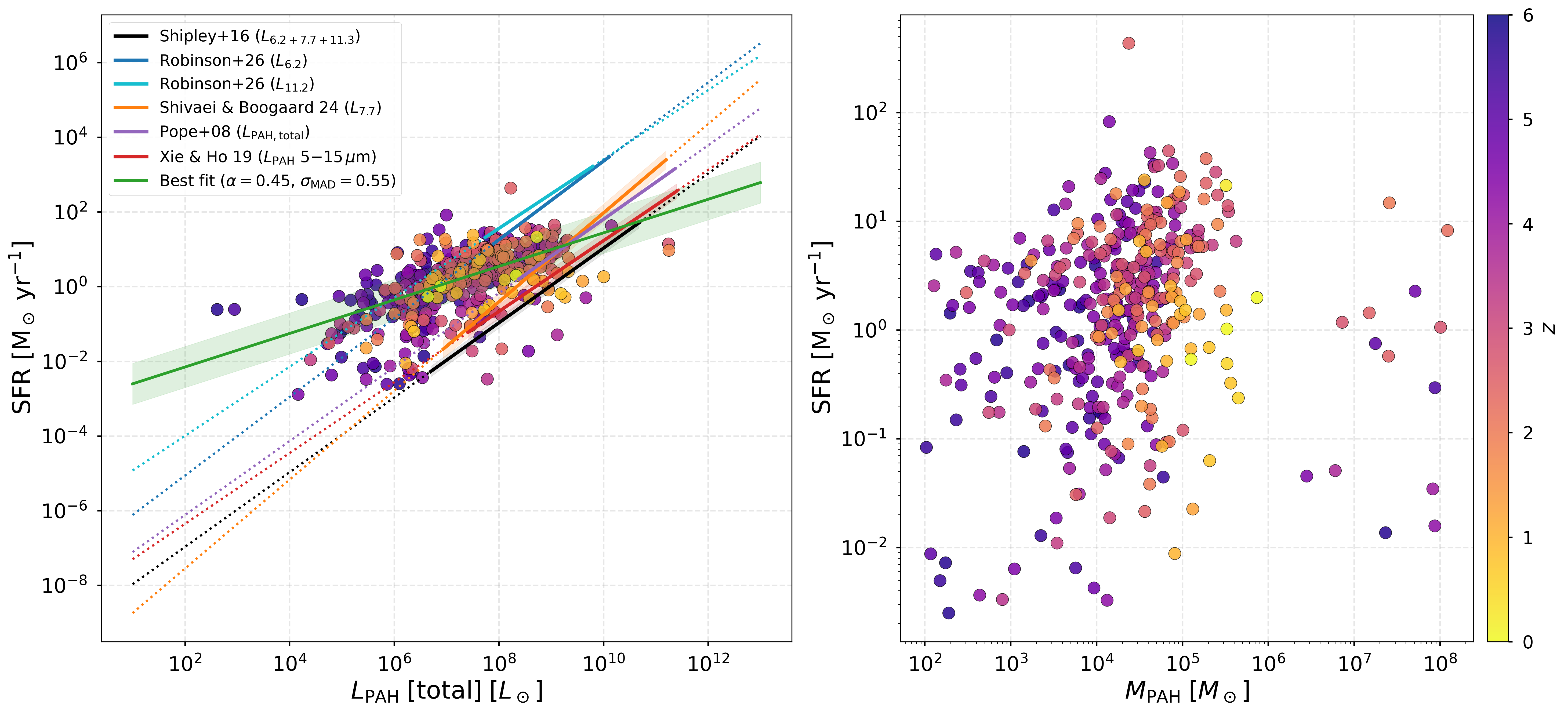}
\caption{{\bf Model SFR vs $L_{\rm PAH}$ (left) and
      SFR vs $M_{\rm PAH}$  (right) relationships, color-coded by galaxy
      redshift.} The individual symbols represent individual galaxy
    snapshots, while the lines represent literature observational
    constraints (with the specific PAH feature observed listed in the
    legend).  The observations come from \citet{shipley16a},
    \citet{robinson26a}, \citet{shivaei24a}, \citet{pope08a}, and
    the $z\approx0$ calibration of \citet{xie19a}; for
    each, the range of constraint provided by observations is shown by
    the solid line, while extrapolations are denoted by dotted
    extensions to the solid line.  The green solid line shows the best
    fit: $\log(\mathrm{SFR}) = 0.45\,\log(L_{\rm PAH}/L_\odot) - 3.05$.  Generally, there is a strong
    correlation between the PAH luminosity and the galaxy SFR.  This
    is due in part to a `bigger things are bigger' correlation
    (i.e. bigger galaxies have higher SFR and higher PAH masses: this
    is demonstrated in the right panel), as well as the fact that
    higher SFR galaxies have, inherently, more luminous PAHs per unit
    PAH mass
    (cf. Figure~\ref{figure:lpah_mpah_sigma_sfr}). \label{figure:lpah_sfr}}
\end{figure*}

\subsection{The PAH-SFR and PAH-CO Relationship in Galaxies}
\label{section:pah_sfr}
We next turn our attention to the $L_{\rm PAH}-$SFR relationship that
has been established observationally in galaxies across a wide dynamic
range in galaxy SFR
\citep{helou04a,smith07a,bendo08a,shipley16a,shivaei24b}.  Indeed,
this has motivated the potential use of PAH features as SFR indicators
in galaxies.  In Figure~\ref{figure:lpah_sfr}, we show the total
$L_{\rm PAH}$ vs SFR for all of our model galaxy snapshots
(color-coded by redshift), compared against published calibrations
from \citet{shipley16a}, \citet{shivaei24b} and \citet{robinson26a},
as well as the $z\approx0$ total (5--15$\mu$m) PAH calibration of
\citet{xie19a}, which anchors the comparison against high-quality
local Spitzer/IRS data.
For each observational constraint, we denote the individual PAH
feature used in the calibration.

Our simulations naturally produce a clear relationship between
  SFR and $L_{\rm PAH}$ that is broadly consistent with the
relatively large dispersion inferred from observational constraints.
The origin of this relationship in our model is twofold.  First, there
is a ``bigger galaxies are bigger'' effect: more massive galaxies have
both higher SFRs and larger dust (and therefore PAH) reservoirs.  We show this in the right panel of
Figure~\ref{figure:lpah_sfr}, which shows the underlying SFR-$M_{\rm
  PAH}$ correlation.  Second, as established in
\S~\ref{section:lpah_mpah}, higher SFR galaxies have elevated
$\Sigma_{\rm SFR}$ and therefore stronger radiation fields, driving a
higher PAH light-to-mass ratio
(Equation~\ref{equation:lpah_mpah_scaling} and
Figure~\ref{figure:lpah_mpah_sigma_sfr}).  The best fit
  relationship to Figure~\ref{figure:lpah_sfr} is $\log(\mathrm{SFR})
  = 0.45\,\log(L_{\rm PAH}/L_\odot) - 3.05$ ($\sigma_{\rm MAD} =
  0.55$~dex), where $\sigma_{\rm MAD}$ is the median absolute
  deviation.  This said, we caution against over-interpreting this
  best fit relationship, as this work investigates cosmological
  zoom-in simulations that have a relatively limited dynamic range in
  physical properties.  Large box cosmological simulations will result
  in more robust theoretical relationships between the galaxy star
  formation rate and $L_{\rm PAH}$, and will be presented in due
  course (M. Parente et al., in prep.). 

\begin{figure*}
\centering
\includegraphics[scale=0.55]{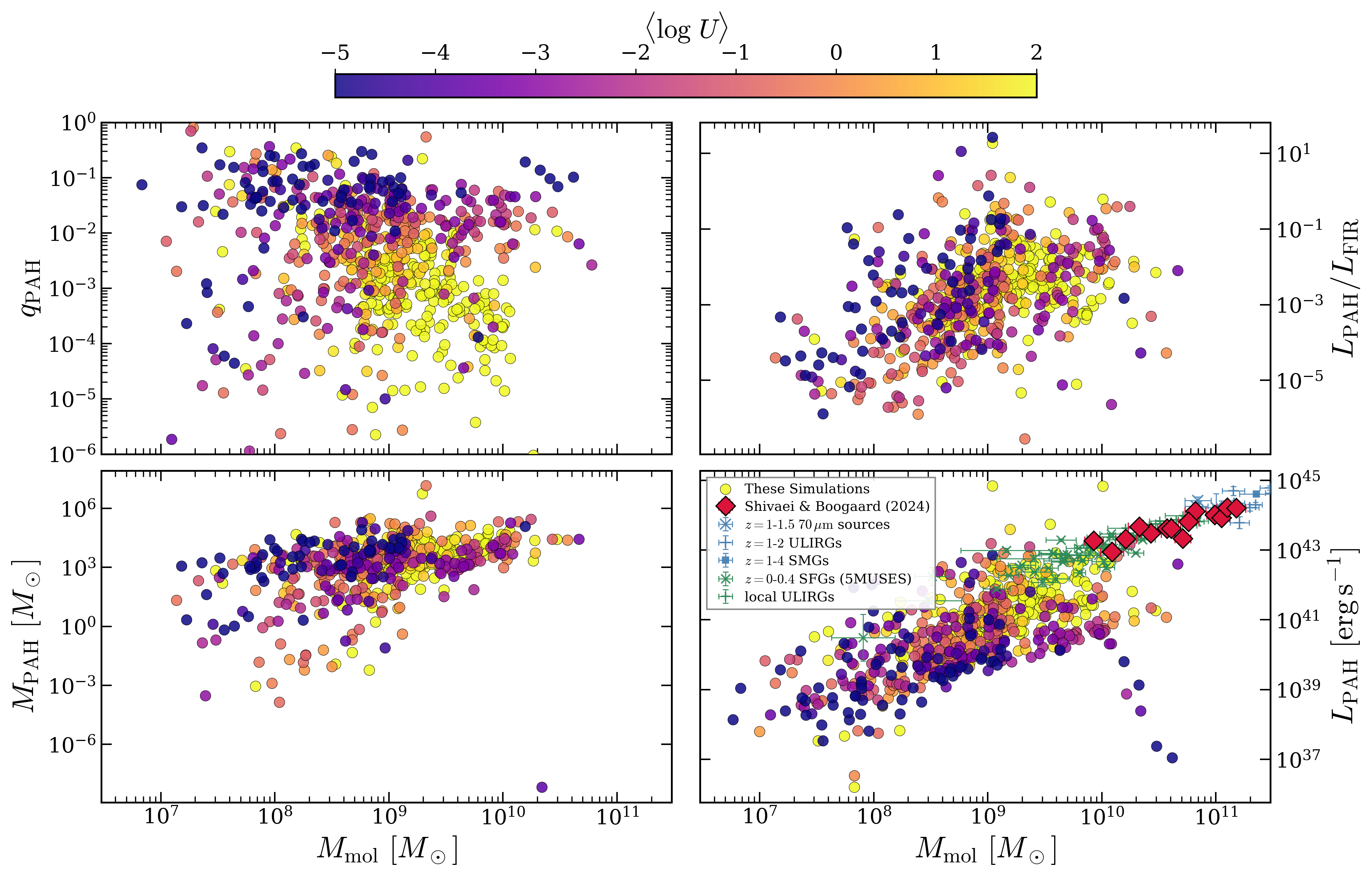}
\caption{{\bf The physical PAH mass fraction $q_{\rm PAH}$ (top left)
    and the observable $L_{\rm PAH}/L_{\rm FIR}$ (top right) as a
    function of molecular gas mass $M_{\rm mol}$, color-coded by
    dust-mass weighted deposited radiation field ($\langle \log\,U
    \rangle$.)}  While $q_{\rm PAH}$ decreases with increasing $M_{\rm
    mol}$ (owing to the fact that grains more efficiently shatter in
  our model in diffuse gas), the observable ratio $L_{\rm PAH}/L_{\rm
    FIR}$ increases with $M_{\rm mol}$.  This discrepancy is driven by
  the interstellar radiation field: the highest molecular gas mass
  galaxies are also the most heavily star-forming, and subsequently
  have the hardest radiation fields (highest $U$), which excite PAHs
  more efficiently per unit PAH mass. In the bottom row we show
  $M_{\rm PAH}$-$M_{\rm mol}$ (left) and $L_{\rm PAH}-M_{\rm mol}$ in
  order to demonstrate reasonable correspondence with observations.
  The observations are taken from \citet{cortzen19a} and
  \citet{shivaei24a}. \label{figure:qpah_mmol}}
\end{figure*}

Building from the intuition gained from Figure~\ref{figure:lpah_sfr},
which demonstrates the impact of the interstellar radiation field on
converting PAH {\it mass}-based scaling relations to PAH {\it
  luminosity}-based scaling relations, we now apply our models to the
inferred PAH-molecular gas relation.  In
Figure~\ref{figure:qpah_mmol}, we plot the relationships between
$q_{\rm PAH}$ and $L_{\rm PAH}/L_{\rm FIR}$ as a function of $M_{\rm
  mol}$ (we additionally show $M_{\rm PAH}-M_{\rm mol}$ and $L_{\rm
  PAH}-M_{\rm mol}$ in order to compare more directly against
observations.  Similar to Figure~\ref{figure:qpah_fmol}, the PAH mass
fraction decreases with the molecular gas mass in galaxies (which has
a weak correlation with $f_{\rm mol}$ in this instance). However, the
observable quantity $L_{\rm PAH}/L_{\rm FIR}$ demonstrates a positive
correlation with $M_{\rm mol}$, meaning that observationally one would
infer an increase in $q_{\rm PAH}$ toward denser phases of gas.  This
point, which we return to in \S~\ref{section:discussion}, is due to
the hardness of the radiation field increasing with $M_{\rm mol}$.  We
parameterize this by color-coding each galaxy snapshot by dust-mass
weighted log$\left(U\right)$.  We remind the reader that $U$ is a
dimensionless quantity parameterizing the rate of energy absorption
onto grains (Equation~\ref{equation:logu}).  The highest molecular gas
mass galaxies are also the most heavily star-forming, and subsequently
have the hardest radiation fields capable of exciting PAHs.  The
upshot is that despite the fact that in our model PAHs decrease in
relative abundance in galaxies dominated by dense molecular gas, they
tend to be {\it brighter} relative to the dust continuum in heavily
star-forming galaxies.  This discrepancy between the physical and
luminous properties of PAHs, driven by the interstellar radiation
field, will be explored further in a forthcoming paper.  We
  emphasize that these trends derive from the PAH formation pathways that we model (i.e., via grain-grain shattering), and the fact that PAHs tend to form in diffuse gas in our model.  This is as opposed to explicit gas-dust-PAH chemistry, which our model does not yet include (\S~\ref{section:caveats}).

%
%
%
%


  

\section{Discussion}
\label{section:discussion}

\subsection{Our Results in the Context of Other PAH Galaxy Evolution Models}

We first place our results in the context of the broader theoretical
landscape for PAH evolution in galaxies.  As far as we know, this
paper represents the first effort to include both the physical and
luminous properties of PAHs in cosmological simulations of any type,
but in particular, hydrodynamic simulations.  Until this point,
PAH-specific models have either focused on one-zone frameworks for
galaxy evolution \citep[e.g.][]{seok14a,hirashita20a,hirashita20b},
idealized galaxy simulations \citep{narayanan23a,rodriguezmontero26a},
or cosmological simulations that include only $2$ size bins for dust
evolution \citep[e.g.][]{aoyama18a,hou19a}.  This presented work aims to fill the gap left by the bespoke studies. 

Our model is developed based on the premise of exploring the ansatz
that PAHs are produced primarily via grain-grain shattering
\citep[i.e., a ``top-down'' formation mechanism][]{narayanan23a}.  We
find that such a model can reasonably reproduce observed PAH mass
fractions at $z=0$, the $q_{\rm PAH}$-metallicity relationship in
galaxies from $z=0-2$, and the $L_{\rm PAH}$-SFR and $L_{\rm
  PAH}-M_{\rm mol}$ relationships in galaxies.  Earlier models
\citep[e.g.][]{seok14a} also relied entirely on a top-down shattering
mode for PAH formation, though this was expanded on by later models
\citep[e.g.][]{hirashita20a}, which added the additional layer of
photo-chemical processing of carbonaceous grains to convert from
aliphatic to aromatic chemical structures.
\citet{rodriguezmontero26a} presented a more complex picture, in which
both top-down shattering as well as bottom-up growth pathways (in
which carbon ions accrete directly onto existing aromatic rings) are
viable pathways for PAH formation.  In this picture, the former
pathway dominates in low metallicity gas, while accretion becomes
efficient in metal-rich, cold and well-shielded gas.  This said,
\citet{choban26a} argue that for bottom-up formation to proceed
without causing runaway growth (i.e., overproducing the $2175 \AA$
bump in local galaxy analogs), there must be some growth limiter in
place for PAHs, such as Coulomb repulsion \citep{weingartner99a}, or
low sticking efficiencies.

On the other hand, \citet{galliano08a}, \citet{bekki13a} and
\citet{murga25a} all suggest that PAH production could be dominated by
direct injection into the ISM by AGB stars.  This said, it is unclear
if PAH formation models in stellar outflows are able to produce
sufficient PAH abundances to match the inferences from local galaxy
constraints \citep{frenklach89a,cherchneff92a}.

These different PAH formation pathways carry distinct implications for
the physical origin of the PZR.  In our model, the PZR arises because
shattering is a collisional process whose rate scales with the
diffuse/dense gas ratio, as well as the abundance of large grains.
Both of these scale with increasing metallicity in our model, driving
a steep rise in \qpah\ with metallicity (\S~\ref{section:pzr}).
\citet{seok14a} arrived at a similar picture using a one-zone chemical
evolution framework.  In their model, shattering of carbonaceous grains
is the sole PAH formation mechanism, and the sharp increase in PAH
abundance above a critical metallicity ($Z \sim 0.1~Z_\odot$) is a
direct consequence of the non-linear dependence of shattering rates on
the total dust-to-gas ratio, which itself undergoes rapid growth once
ISM accretion overtakes stellar dust production
\citep[e.g.][]{narayanan26a}.  Observationally, \citealt{tarantino25a}
have recently detected PAH emission in compact clumps in the dwarf
galaxy Sextans~A at $\sim 0.07~Z_\odot$, near this critical threshold.
\citet{rodriguezmontero26a} present a more nuanced picture in which
the PZR could be driven by two distinct channels operating in
different metallicity regimes (i.e. shattering in the low-metallicity
regime, and growth in the high-metallicity regime).  In contrast to
these models, the stellar seeding based models of \citet{galliano08a}
and \citet{bekki13a} attribute the PZR to the delayed injection of
carbonaceous dust by AGB stars: because carbon-rich AGB stars require
$\sim$a few hundred Myr to evolve off the main sequence, young
low-metallicity systems have not yet had time to build a significant
PAH reservoir.  In this interpretation, the PZR is fundamentally an
age-metallicity effect rather than an ISM processing effect.


\subsection{Relationship between our Model and Observations}
JWST has enabled spatially resolved observations of PAHs in nearby
galaxies.  While we have focused our model comparisons against
globally integrated observations, we now speculate on the relationship
between our presented model, and inferences from spatially resolved
observations.  We note that a full investigation of the relationship
between our model for the lifecycle of PAHs and observations {\it
  within} galaxies will be presented in a future work.

Our model for the lifecycle of PAHs in cosmologically
evolving galaxies results in the primary PAH formation channel being
through grain-grain shattering of larger grains; this process is
facilitated by the presence of diffuse gas in the ISM, where
collisional velocities between grains may be higher.  While our model
reasonably reproduces the globally integrated PZR as measured from
$z=0-2$ (cf. Figure~\ref{figure:pzr}), recent high-resolution
observations by JWST of M101 have evidenced a {\it spatially resolved}
PZR \citep{whitcomb25a}.  Below a threshold metallicity of $Z_{\rm
  th} \approx 0.6\,Z_\odot$, \citet{whitcomb25a} find that
$\Sigma_{\rm PAH}/{\rm TIR}$ drops rapidly for all PAH features,
while the fractional contribution of the 3.3~$\mu$m feature to total
PAH luminosity actually {\it increases} by a factor of $\sim 3$
between $Z_\odot$ and $0.4\,Z_\odot$.  It is not entirely clear how a
picture of a spatially resolved PZR aligns with the physical
underpinnings of the model that we present here.  One possibility is that in
lower metallicity gas within galaxies, grain growth has not been
sufficient to produce significant numbers of large grains to collide
and shatter.  Indeed, \citet{narayanan26a} found that the density of
nearby metals was a strong indicator of grain growth in similar
simulations to those presented here.  On the other hand,
\citet{whitcomb25a} advocate for a scenario in which the growth of
small PAHs {\it into} larger PAHs  is inhibited at
low metallicity, shifting the overall size distribution toward smaller
grains.  While both mechanisms are related to grain growth, the two
scenarios address the potential existence of a spatially resolved PZR
from different ends of the size spectrum: our model would require
either lower interstellar collision velocities, or a lack of large
grains to shatter in lower metallicity gas within galaxies.
This is in contrast to the \citet{whitcomb25a} scenario, which posits
that PAHs form from the bottom up but cannot efficiently grow into
larger sizes at low metallicity.

A related question concerns the observed relationship between PAH
emission, inferred mass fraction, and ISM phase in nearby
galaxies. For example, resolved observations of PAH emission in the
Magellanic clouds show increased PAH emission in molecular dominated
regions \citep{sandstrom12a,chastenet19a}, a result confirmed by
\citet{tarantino25a} in Sextans A.  Within the Galaxy,
\citet{hensley22b} find that the fraction of cold neutral medium is
tightly correlated with $q_{\rm PAH}$.  While our models demonstrate
that on global, unresolved scales, the fraction of dust emitting in
PAHs increases with the molecular gas mass in galaxies, it is not
entirely clear whether or not our hypothesis that PAHs originate in a
top-down shattering-dominated picture is in agreement with these
resolved observations.  Worse yet is that our arguments for why PAH
luminosity scales with molecular gas mass (or star formation rate) is
difficult to falsify, given the underlying premise of ``PAHs are there
in low star formation rate[molecular gas mass] galaxies, but you just
cannot see them.''  Measurements of variations in the dust extinction
law as a function of ISM phase (and, in particular, the $2175$ \AA
\ UV bump, if it indeed is connected to PAHs) may help to provide
evidence in support of, or against these presented models.  Beyond
this, we encourage continued mapping of PAH emission from a diverse
range of ISM phases near and far.  Indeed, at least some observations
of nearby galaxies suggest that a significant fraction of PAH emission
arises in the diffuse ISM \citep{evans22a}.

\subsection{Caveats and Missing Physics in our Modeling}
\label{section:caveats}
Finally, we discuss a number of points of improvement in our modeling
technique, and room for future study.  There are four major
uncertainties in our model that may impact the interpretation of our
results.  While each was discussed in context throughout the paper, we
consolidate them in this section for the reader's convenience.

The most fundamental caveat in our model is our treatment of PAHs
  as the smallest carbonaceous grains in the size distribution, as
  opposed to molecules.  This approximation has two principal
  consequences.  First, in our model PAHs are formed through the same
  collisional network as the rest of the dust grain population (i.e.,
  shattering), while coagulation can remove PAHs in dense gas.
  Whether PAHs as molecules are actually processed in a top-down
  channel like this (as opposed to, or potentially in addition to
  bottom-up growth through gas-phase and grain-surface chemistry) is
  not established in the literature.  Second, we do not model
  processes that operate uniquely on molecules: photo-ionization and
  photo-dissociation by the interstellar radiation field, and
  molecular chemical reactions \citep[e.g.][]{tielens08a}.  At the
  same time, the molecular nature of PAHs is not entirely absent from
  our framework: the luminous properties of PAHs are computed in the
  single-photon limit, in which each absorbed UV photon stochastically
  excites an individual PAH molecule that subsequently cools by
  radiating in the classical mid-IR bands \citep{richie25a}.  In this
  sense our model is a hybrid, including a grain-like lifecycle,
  coupled with molecular emission physics.  The agreement of our model
  with the majority of observed scaling relations suggests that this
  remains an adequate effective description of PAH physics at the
  galaxy population level, though we caution that this success does
  not by itself validate the underlying microphysics.  An explicitly
  molecular treatment that tracks distinct chemical species, ring
  structure, and growth and destruction via molecular chemical
  reaction networks, all coupled with radiative transfer remains a
  computational challenge, and is a clear direction for future models in this class.

Second, we consider all grains with sizes $a \la 13 \AA$ that are
carbonaceous dominated to be PAHs.  This corresponds to roughly $\sim
1000$ carbon atoms \citep{hensley23a}, which is an approximate upper
size limit for PAHs \citep{draine21a}.  What this is missing is a
chemical differentiation between aliphatic and aromatic ultrasmall
carbonaceous grains.  While aliphatic grains are of course evident in
the mid-IR spectra of galaxies, we do not consider them in this model.
Aromatization may involve dehydrogenation in strong radiation fields
\citep[e.g.][]{murga19a}, while the physical conditions associated
with structural changes in chemical bonds that drive aliphatization
are not well constrained.  Including a robust model for the chemical
structure of ultrasmall carbonaceous grains would represent a
significant improvement to the model here.

Similarly, there are two major potential impacts from active galactic
nuclei (AGN) on PAHs that are not considered here.  The first is a
swamping of the mid-IR emission from AGN continuum
\citep[e.g.][]{hopkins07a,nenkova08a}.  Future work will include AGN
in our radiative transfer models \citep[e.g.][]{narayanan21a} to
investigate this.  Second, we do not include the putative destruction
of PAHs in the strong radiation fields surrounding AGN.  Incorporation
of radiative destruction of PAHs both in the vicinity of AGN, as well
as in H~{\sc ii} regions \citep{egorov23a,sutter24a} constitutes an
important potential improvement to this model.

\section{Summary}
\label{section:summary}

In this paper, we have presented the first cosmological hydrodynamic
simulations designed to model both the physical lifecycle and luminous
properties of polycyclic aromatic hydrocarbons (PAHs) in galaxies as
they evolve from $z=6\rightarrow 0$.  We have combined $40$
cosmological zoom-in simulations with an on-the-fly
model for the formation, growth, and destruction of multi-species,
multi-size dust grains, and coupled these with single-photon excitation
calculations and dust radiative transfer to compute the emergent
mid-infrared spectra.  We operate under the ansatz that PAHs are
ultrasmall ($a < 13$~\AA) carbonaceous dust grains that form
predominantly via grain-grain shattering, and investigate the physical
processes governing their abundance and luminosity across cosmic time.
Our main results follow.

\begin{enumerate}

\item {\bf The dominant physical driver of PAH formation in our
  simulations is grain-grain shattering in diffuse ISM gas.}  We
  derive the scaling of the dust grain interaction velocity with gas
  properties (Equation~\ref{equation:vb_rhot}) and demonstrate that the local gas
  density is the single most important factor controlling shattering
  rates: low-density, diffuse gas decouples dust grains from the
  turbulent flow, enabling large relative velocities.  As galaxies
  evolve, the typical ISM conditions surrounding dust shift from dense
  and molecular at high redshift to increasingly diffuse at low
  redshift (Figure~\ref{figure:dust_density_temp_z}), resulting in a
  corresponding rise in grain-grain collision velocities
  (Figure~\ref{figure:vdisp_z}) and a transfer of power from large
  grains to ultrasmall grains in the grain size distribution
  (Figure~\ref{figure:gsd_z}).  The consequence is a rise in the PAH
  mass fraction from $q_{\rm PAH} \sim 5 \times 10^{-4}$ at $z \sim 4$ to $\sim
  10^{-2}$ at $z \sim 0$ (Figure~\ref{figure:qpah_ridge}), with a
  corresponding morphological evolution from compact, sparse PAH
  distributions at high redshift to extended structures tracing
  diffuse spiral arms at $z = 0$
  (Figure~\ref{figure:pah_morphology_evolution}).

\item {\bf The PAH mass fraction anti-correlates with the molecular gas
  fraction in our models.}  Increased PAH production in diffuse gas
  results in an inverse relationship between $q_{\rm PAH}$ and the
  molecular gas fraction $f_{\rm mol}$: when $f_{\rm mol}$ is low, dust
  grains reside preferentially in diffuse environments where collision
  velocities are high and shattering is efficient, driving $q_{\rm PAH}$
  upward (Figure~\ref{figure:qpah_fmol}).

\item {\bf Among the destruction pathways explored, thermal sputtering
  is the dominant mechanism for destroying PAHs in our simulations,
  though its impact is
  modest (Figure~\ref{figure:mpah_z_destruction}).}  This said, we
  caution that our models do not include the radiative dissociation of
  small grains or destruction near H~{\sc ii} regions, both of which
  may be important.

\item {\bf The PAH light-to-mass ratio scales linearly with the
  radiation field intensity ($L_{\rm PAH}/M_{\rm PAH} \propto G_0$),
  but anti-correlates with $q_{\rm PAH}$.}  When PAH emission
  operates in the single-photon regime, the luminosity per unit PAH
  mass traces the absorbed photon rate rather than a grain temperature
  (Equation~\ref{equation:lpah_mpah_scaling} and
  Figure~\ref{figure:lpah_mpah_sigma_sfr}).  At the same time, high
  $\Sigma_{\rm SFR}$ galaxies tend to have higher molecular gas
  fractions and therefore lower shattering rates, resulting in an
  inverse relationship between $L_{\rm PAH}/M_{\rm PAH}$ and $q_{\rm
    PAH}$.  This decoupling means that the physical $q_{\rm PAH}$ and
  the observed $L_{\rm PAH}/L_{\rm FIR}$ do not evolve in lockstep
  (Figures~\ref{figure:sed_gallery} and ~\ref{figure:qpah_z}).

\item {\bf The PAH-metallicity relationship (PZR) arises naturally in
  our framework.}  As galaxies enrich over cosmic time, they
  simultaneously grow their diffuse ISM mass fraction, linking rising
  metallicity to rising $q_{\rm PAH}$ via increased grain-grain
  shattering rates.  The observed PZR additionally encodes the
  evolving radiation field conditions that modulate the PAH
  light-to-mass ratio.  Our models represent the first to reasonably reproduce the
  PZR observed across $z = 0$-$2$
  (Figures~\ref{figure:pzr_lpah_lfir_z_Z}
  and~\ref{figure:pzr}).

\item {\bf A relationship between $L_{\rm PAH}$ and SFR  emerges from the
  combination of two effects.}  First, more massive galaxies have both
  higher SFRs and larger PAH reservoirs, producing a ``bigger things
  are bigger'' correlation.  Second, higher-SFR galaxies have elevated
  $\Sigma_{\rm SFR}$ and therefore stronger radiation fields, driving a
  higher PAH light-to-mass ratio
  (Figure~\ref{figure:lpah_sfr}).

\item {\bf While $q_{\rm PAH}$ decreases with the molecular gas
    fraction in our models (Figure~\ref{figure:qpah_fmol}), the
    observable $L_{\rm PAH}/L_{\rm FIR}$ {\it increases} with galaxy
    molecular gas mass.}  This owes to harder radiation fields in more
    heavily star-forming galaxies, which tend to have larger molecular
    gas reservoirs (Figure~\ref{figure:qpah_mmol}).
  
\end{enumerate}

\section{Acknowledgments}
D.N. expresses gratitude toward the Aspen Center for Physics, which is
supported by National Science Foundation grant PHY-1607611, where the
idea for this series of projects was borne out on repeated chair-lift
rides during a fantastic day on the slopes with J.D.S.  The authors
are grateful to NASA for funding this study, via grants ATP-21-0013
(PI: Torrey \& Narayanan), ATP-23-0002 (PI: Narayanan \& Torrey), and
ADSPS-23-0007 (PI: Narayanan, Smith, Pope).  This research was carried
out in part at the Jet Propulsion Laboratory, California Institute of
Technology, under a contract with the National Aeronautics and Space
Administration (80NM0018D0004).  IS acknowledges funding from the
European Research Council (ERC) DistantDust (Grant No.101117541) and
the Atracc\'{i}on de Talento Grant No.2022-T1/TIC-20472 of the
Comunidad de Madrid, Spain. The authors acknowledge UFIT Research
Computing for providing computational resources on HiPerGator and
support that have contributed to the research results reported in this
publication (\url{http://it.ufl.edu/rc}).  The authours used
Anthropic's Opus models for plot aesthetics; all code was reviewed by D.N.


\bibliographystyle{mnras}
\bibliography{full_refs,extra_refs}

\clearpage

\appendix

\section{Galaxy Physical Properties}
\label{appendix:galaxy_physical_properties}
\begin{figure*}[!b]
   \centering
   \includegraphics[scale=0.4]{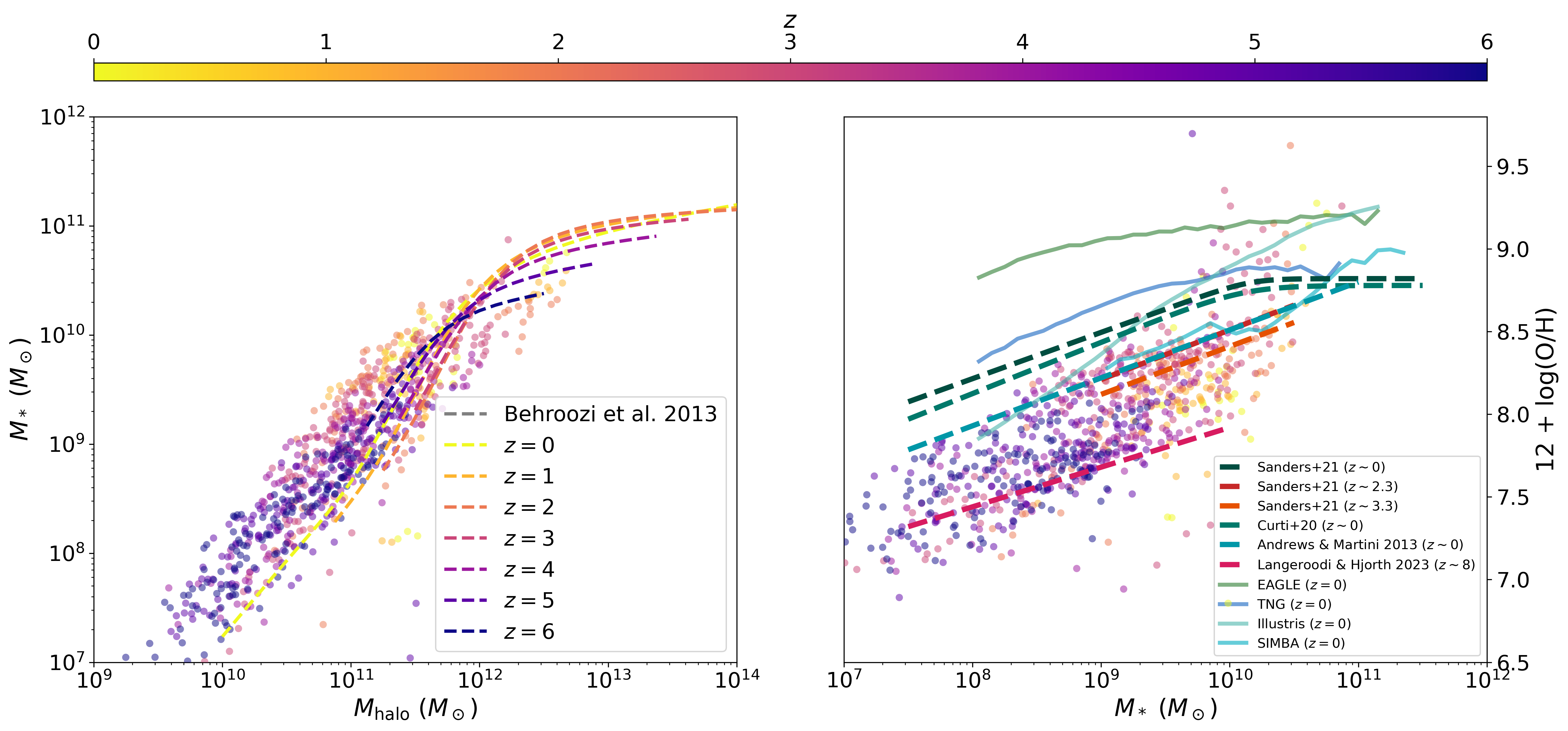}
   \caption{{\bf A comparison of simulated galaxies to $M_*-M_{\rm halo}$ relationship (left) and $M_*-Z_{\rm gas}$ relationship (MZR; right) shows reasonable correspondence.}  Each point in both panels shows an individual galaxy snapshot.  {\it Left:} The stellar mass-halo mass relationship for our model galaxies (circles) is compared to the \citet{behroozi13a} inferences when operating under the abundance matching ansatz (dashed lines).  Colors show redshift of both simulations and observations.     {\it Right:} The observational constraints for the MZR  are from  \citet{andrews13a},
     \citet{curti20a}, \citet{langeroodi23a} and \citet{sanders21a} are shown as dashed curves, respectively, while the $z=0$
     predictions from the {\sc eagle} , {\sc tng}, {\sc illustris},
     and {\sc simba} cosmological simulations are shown as solid curves \citep[compiled by][]{garcia24a}.  Our model
     galaxies enrich with decreasing redshift and lie reasonably
     within the locus of observational and theoretical constraints. \label{figure:behroozi}}
 \end{figure*}

We verify the core physical properties of our model galaxies against
standard observational constraints for cosmological simulations: the
$M_*-M_{\rm halo}$ relationship, and the $M_*-Z_{\rm gas}$
relationship (hereafter, the MZR) in Figure~\ref{figure:behroozi}.  For the former, we compare against constraints
from \citet{behroozi13a}, while for the latter we compare against
observational inferences at a range of redshifts
\citep{andrews13a,curti20a,sanders21a}, as well as the $z\sim0$
results from major cosmological simulations as compiled by
\citet{garcia24a}.  Our simulations demonstrate reasonable
correspondence with observations and/or other simulations for both relationships.

\begin{figure*}
   \centering
   \includegraphics[scale=0.4]{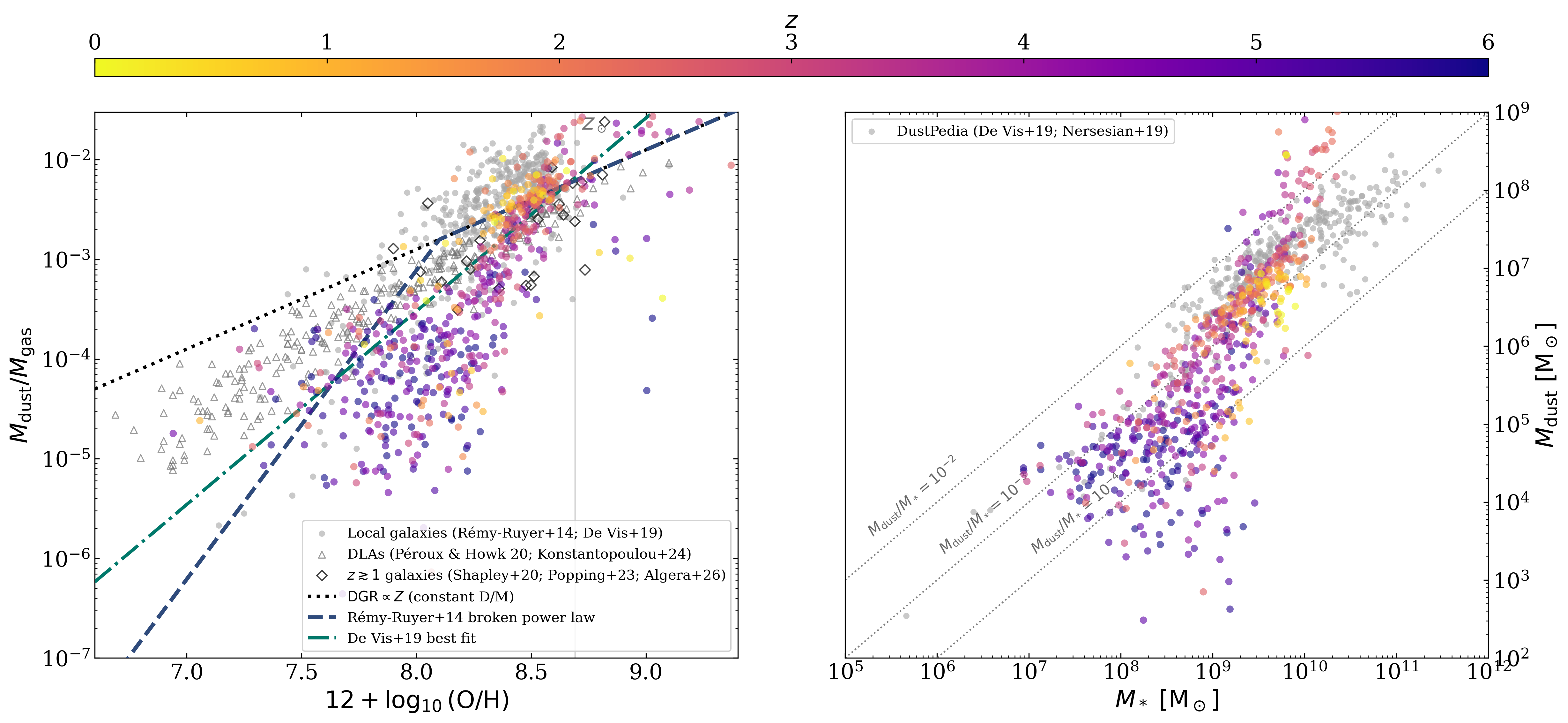}
   \caption{{\bf The dust scaling relations of our model galaxies:
     dust-to-gas ratio vs. metallicity (left) and dust mass vs. stellar
     mass (right).}   Each circle shows an individual
       galaxy snapshot, colored by redshift.  {\it Left:} Gray points show observational
     constraints from local galaxy surveys
     \citep{remyruyer14a,devis19a}, damped Lyman-$\alpha$ absorbers
     \citep{peroux20a,konstantopoulou24a}, and emission-based
     measurements of $z \gtrsim 1$ galaxies
     \citep{shapley20a,popping23a,heintz25a,algera26a}.  The dotted line
     shows a linear ${\rm DGR} \propto Z$ scaling at a fixed
     dust-to-metals ratio (normalized to the solar dust-to-gas ratio),
     while the dashed and dash-dotted lines show the
     \citet{remyruyer14a} broken power-law and \citet{devis19a} best
     fits to local galaxies.  Our galaxies lie below the local relations
     at early times and converge onto them with decreasing redshift as
     accretion-driven grain growth comes to dominate the dust mass.
     {\it Right:} Gray points show the DustPedia nearby-galaxy sample,
     with dust masses from \citet{devis19a} and stellar masses from
     \citet{nersesian19a}; dotted lines mark constant
     $M_{\rm dust}/M_* = 10^{-4}$, $10^{-3}$, and
     $10^{-2}$. \label{figure:dgr_z}}
 \end{figure*}

In Figure~\ref{figure:dgr_z}, we additionally show the dust-to-gas
  ratio of our model galaxies as a function of gas-phase metallicity
  (left), as well as the dust mass--stellar mass relation (right).  At
  fixed metallicity, the dust-to-gas ratio of our galaxies rises with
  decreasing redshift, converging onto the observed $z\sim0$ relations
  as ISM growth overtakes stellar production in the dust budget.
  Similarly, our galaxies build toward specific dust masses $M_{\rm
    dust}/M_* \sim 10^{-3}-10^{-2}$ by late times, overlapping the
  observed locus of the DustPedia nearby-galaxy sample
  \citep{devis19a,nersesian19a}.

\section{Derivation of $\Lpah/\Mpah \propto \Gzero$}
\label{appendix:lpah_derivation}

We derive the total power emitted by a single PAH by integrating
equation~(\ref{eq:normalized_spectrum}) over all emission wavelengths and summing
over all absorbed-photon channels:
\begin{align}
  \ell_{\rm PAH}
  &\;\equiv\;
  \int p_{\lambda_{\rm em}}\,d\lambda_{\rm em}
  \;=\;
  \sum_{\lambda_{\rm abs}}
  \int p_{\lambda_{\rm em}}(\lambda_{\rm abs})\,d\lambda_{\rm em}
  \notag\\
  \intertext{Substituting equation~(\ref{eq:normalized_spectrum}) and integrating
    over $\lambda_{\rm em}$, the $\tilde{p}$ integrals in the numerator and
    denominator cancel, leaving just the absorbed power in each wavelength bin:}
  &\;=\;
  \sum_{\lambda_{\rm abs}}
  \int_{\lambda_{\rm abs}}^{\lambda_{\rm abs}(1+\Delta)}
  c\,u_\lambda(\lambda)\,C_{\rm abs}(\lambda)\,d\lambda
  \notag\\
  \intertext{Summing over bins reassembles this into the full integral over the
    UV/optical spectrum.  Using $C_{\rm abs} \approx \pi a^2 \Qabs^{\rm UV}$ and
    $u_\lambda \propto \Gzero$:}
  &\;=\;
  \int C_{\rm abs}(\lambda)\,c\,u_\lambda\,d\lambda
  \;\propto\;
  \pi a^{2}\,\Qabs^{\rm UV}\,\Gzero .
  \label{eq:lpah_grain_appendix}
\end{align}
Normalizing by grain mass $m_{\rm gr} = \tfrac{4}{3}\pi a^3 \rho_{\rm gr}$, with
$N$ canceling between $\Lpah = N\,\ell_{\rm PAH}$ and $\Mpah = N\,m_{\rm gr}$:
\begin{equation}
  \frac{\Lpah}{\Mpah}
  \;\propto\;
  \frac{\pi a^2\,\Qabs^{\rm UV}}{\tfrac{4}{3}\pi a^3\,\rho_{\rm gr}}\,\Gzero
  \;=\;
  \underbrace{\frac{3\,\Qabs^{\rm UV}}{4\,a\,\rho_{\rm gr}}}_{\displaystyle\equiv\,\Kappa_{\rm UV}^{\rm PAH}}
  \Gzero,
\end{equation}
where $\Kappa_{\rm UV}^{\rm PAH}$ (cm$^2$\,g$^{-1}$) is the UV opacity per unit PAH
mass.  For a realistic size distribution $\Kappa_{\rm UV}^{\rm PAH}$ becomes a
size-weighted average, though remains linearly dependent on $\Gzero$.

{
\section{Details of the {\sc powderday} Dust Radiative Transfer}
\label{appendix:rt_details}

Here we expand on the radiative transfer methodology summarized in
\S~\ref{section:powderday}; complete descriptions are given in the
{\sc powderday} code paper \citep{narayanan21a}, the {\sc hyperion}
code paper \citep{robitaille11a}, and \citet{narayanan23a}.

{\it Grid construction:} We build the radiative transfer grid directly
on a Voronoi tessellated grid, built on the dust particles themselves.
Each Voronoi cell inherits the dust mass, density, grain size
distribution and composition.

{\it Photon transport:} {\sc hyperion} solves the
transfer problem via Monte Carlo photon packet propagation.  Here, photons are emitted isotropically from star particles, and each packet is assigned a 
 random optical depth $\tau = -\ln \xi$ (with $\xi$ a
uniform random deviate) to determine when it will next interact. Photons are 
propagated cell by cell through the tessellation: within a cell the
density and extinction properties are uniform, so the geometric path
length $\Delta s$ through the cell (computed exactly from the
intersection of the packet trajectory with the cell faces) converts
directly to an optical depth increment $\Delta \tau = \rho \,
\kappa_{\rm ext}(\lambda) \, \Delta s$, with $\kappa_{\rm
  ext}(\lambda)$ set by the local grain size distribution and
composition.  When the accumulated optical depth reaches the sampled
value, the packet scatters or is absorbed and re-emitted according to
the local albedo.  The mean radiation field in each cell is computed
from the \citet{lucy99a} path-length estimator, which accumulates the
energy deposited by every packet traversal of the cell, and the
transfer is iterated until the absorbed energy in each cell converges
to better than 1\%.

{\it Emergent SEDs:} Once the radiation field has converged and the
PAH emission spectra have been added as source terms (with the
transfer iterated again to capture attenuation of the PAH features
themselves), the emergent SED is computed by ray tracing: the source
function is evaluated in every cell, and the formal solution of the
radiative transfer equation is integrated through the Voronoi grid
along the line of sight to the virtual observer, with scattered light
accumulated via the peeling-off technique described in
\S~\ref{section:powderday}.}

\end{document}